\documentclass{LMCS}

\usepackage{amssymb}
\usepackage{amsmath}
\usepackage{latexsym}
\usepackage{leqno}
\usepackage{hyperref}

\newcommand{\SQEMA}{\textsf{SQEMA}}

\newcommand{\dia}{\Diamond}

\renewcommand{\phi}{\varphi} 

\newcommand{\nomi}{\mathbf{i}}
\newcommand{\nomj}{\mathbf{j}}
\newcommand{\nomk}{\mathbf{k}}

\newcommand{\forces}{\Vdash}                

\newcommand{\linkerhaak}{[ \! [}
\newcommand{\rechterhaak}{] \! ]}

\newcommand{\lra}{\leftrightarrow}                

\newcommand{\ra}{\rightarrow}

\newcommand{\lbrak}{[}
\newcommand{\rbrak}{]}

\def\doi{2 (1:5) 2006}
\lmcsheading%
{\doi}
{26}
{}
{}
{Sep.~28, 2005}
{Mar.~\phantom{0}8, 2006}
{}
\begin{document}

\title[Algorithmic correspondence and completeness in modal
logic]{Algorithmic correspondence and completeness in modal logic.
\\ I. The core algorithm {\Large SQEMA}}

\author[W.~Conradie]{Willem Conradie\rsuper a}
\address{{\lsuper a}Department of Mathematics, University of Johannesburg, and \vspace{-2mm}}
\address{School of Mathematics, University of the Witwatersrand, Johannesburg}
\email{wec@rau.ac.za}

\author[V.~Goranko]{Valentin Goranko\rsuper b}
\address{{\lsuper b}School of Mathematics, University of the Witwatersrand, Johannesburg}
\email{goranko@maths.wits.ac.za}

\author[D.~Vakarelov]{Dimiter Vakarelov\rsuper c}
\address{{\lsuper c}Faculty of Mathematics and Computer Science, Sofia University}
\email{dvak@fmi.uni-sofia.bg}


\begin{abstract}
  Modal formulae express monadic second-order properties on Kripke
frames, but in many important cases these have first-order
equivalents. Computing such equivalents is important for both logical
and computational reasons. On the other hand, canonicity of modal
formulae is important, too, because it implies frame-completeness of
logics axiomatized with canonical formulae.

   Computing a first-order equivalent of a modal formula amounts to
elimination of second-order quantifiers. Two algorithms have been
developed for second-order quantifier elimination: SCAN, based on
constraint resolution, and DLS, based on a logical equivalence
established by Ackermann.

   In this paper we introduce a new algorithm, SQEMA, for computing
first-order equivalents (using a modal version of Ackermann's lemma)
and, moreover, for proving canonicity of modal formulae. Unlike SCAN
and DLS, it works directly on modal formulae, thus avoiding
Skolemization and the subsequent problem of unskolemization. We
present the core algorithm and illustrate it with some examples. We
then prove its correctness and the canonicity of all formulae on which
the algorithm succeeds. We show that it succeeds not only on all
Sahlqvist formulae, but also on the larger class of inductive
formulae, introduced in our earlier papers. Thus, we develop a purely
algorithmic approach to proving canonical completeness in modal logic
and, in particular, establish one of the most general completeness
results in modal logic so far.
\end{abstract}

\maketitle

\vskip-\bigskipamount
\section*{Introduction}
The correspondence between modal logic and first-order logic has been studied extensively, see for example
\cite{vanBenthemBook}. Every modal formula defines a (local or global) second-order condition on frames, but it
is well known that many of these can in fact be equivalently reduced to first-order conditions. Modal formulae
for which this is possible are called \emph{elementary}. As Chagrova has shown in \cite{Chagrova}, the class of
elementary modal formulae is undecidable. Hence, any attempt at an effective characterization of this class can
be an approximation at best.

The best-known syntactic approximation is the class of \emph{Sahlqvist formulae} \cite{Sahlqvist}, the
first-order correspondents of which can be computed using the Sahlqvist-van Benthem method of substitutions, see
e.g. \cite{vanBenthemBook} and \cite{MLBook}.

The Sahlqvist formulae are properly included in the (still syntactically defined) \emph{inductive formulae},
introduced in \cite{GorankoVakarelov1} and \cite{GorankoVakarelov3}; for a survey on these, see also
\cite{CGVAiml5}. The \emph{complex formulae} (\cite{Vakarelov1}) represent yet another extension of the
Sahlqvist-class. The formulae in these three classes have the important property of being \emph{canonical}, and
hence of axiomatizing complete logics.

Computing first-order equivalents of modal formulae amounts to \emph{elimination of the second-order
quantifiers} in their semantic translation. Thus, algorithmic classes of elementary modal formulae can be
generated by algorithms for elimination of second-order predicate quantifiers, such as SCAN (see
\cite{GabbayOlbach,Engel,Szalas2}) and DLS (first introduced in \cite{Szalas1} as an extension of an algorithm
presented in \cite{Szalas3}, see also \cite{Szalas1,Szalas2,Szalas4} and \cite{NonnengartSzalas}). Each of them,
applied to the standard translation into second-order logic of a negated modal formula, attempts to eliminate
the existentially quantified predicate variables in it and thus to compute a first-order correspondent. Both
SCAN and DLS have been applied to compute the first-order equivalents of standard translations of modal
formulae, and can be regarded as the first purely algorithmic approaches to the first-order correspondence
theory of modal formulae. \cite{GabbayOlbach} and \cite{Szalas3} were soon followed by Simmons' algorithm
presented in \cite{Simmons} as an extension of Sahlqvist-van Benthem method of substitutions (for the Simmons
algorithm see also \cite{Rij93}). Simmons' algorithm is also applicable to some non-elementary modal formulae
(e.g. modal reduction schemes such as McKinsey's formula), but the output formulae produced by it generally
involve Skolem functions.

Briefly, SCAN accomplishes its task on an existential second-order input formula by removing the existential
second-order quantifiers, skolemizing away the existential first-order quantifiers, and transforming the result
into clausal form. It then attempts to obtain an equivalent formula in which the predicate variables do not
occur anymore, by running a constraint resolution procedure to generate sufficiently many first-order
consequences and by applying deletion rules. The introduced Skolem functions are then eliminated from the
result, if possible. SCAN fails to find a first-order equivalent either if the constraint resolution phase does
not terminate, or if it cannot unskolemize. It has been proved in \cite{SCANComplete} that SCAN can successfully
compute the first-order equivalents of all Sahlqvist formulae.

The algorithm DLS is based on the following lemma by Ackermann, see \cite{Ackermann}.

\begin{lem}[Ackermann's Lemma]
Let $P$ be a predicate variable and $A(\overline{x}, \overline{z})$ and $B(P)$ be first-order formulae such that
there are no occurrences of $P$ in $A(\overline{x}, \overline{z})$. If $P$ occurs only negatively in $B(P)$ then
$$\exists P \left( \forall \overline{x}
[A(\overline{x},\overline{z}) \ra P(\overline{x})] \land B(P)
\right) \equiv B( A(\overline{t},\overline{z})/ P(\overline{t}))$$
and, respectively, if $P$ occurs only positively in $B(P)$, then
$$\exists P \left( \forall \overline{x} [ P(\overline{x}) \ra
A(\overline{x},\overline{z})] \land B(P) \right) \equiv B( A(\overline{t},\overline{z}) / P(\overline{t}))$$
where $\overline{z}$ are parameters, and each occurrence of $P(\overline{t})$ in $B$ on the right hand side of
the equivalences, for terms $\overline{t}$,  is substituted by $A(\overline{t},\overline{z})$.
\label{AckermannLemma}
\end{lem}

The algorithm DLS performs various syntactic manipulations, including skolemization, on its input in order to
transform it into a form suitable for the application of Ackermann's lemma. If that cannot be accomplished, the
algorithm fails. When all predicate variables have been eliminated it may be necessary to unskolemize the
resulting formula in order to obtain a first-order equivalent. If it cannot do that, the algorithm fails.

Both SCAN and DLS have been implemented and are available online. For in-depth discussion of the theory and
implementation details of these algorithms see \cite{Engel} (for SCAN) and \cite{Gustafsson} (for DLS).

\medskip

In this paper we present a new algorithm called \SQEMA\ ({\bf
S}econd-Order {\bf Q}uantifier {\bf E}limination for {\bf M}odal
formulae using {\bf A}ckermann's lemma) for computing the
first-order frame correspondents of modal formulae. It is
influenced by both SCAN and DLS, but unlike them it works directly
on modal formulae, rather than on their standard translation, and
computes their (local) first-order equivalents by reducing them to
(locally) equivalent pure formulae in a hybrid language (see e.g.
\cite{MLBook} or \cite{GorankoVakarelov2}) with nominals and
inverse modalities. This is done using a modal version of
Ackermann's lemma. This approach originated from \cite{Vakarelov2}
(see also \cite{Vakarelov3}), where Ackermann's lemma was
rephrased in the framework of solving equations in Boolean and
modal algebras. The fact that modal formulae are not translated
has many advantages, e.g.:

\begin{itemize}
\item it avoids the introduction of Skolem functions, and hence the need for unskolemization, which is in general
undecidable. Nominals play the role of Skolem constants, but disappear when translating the resulting pure
hybrid formulae into first-order logic.

\item it enables the use of a stronger version of Ackermann's
lemma, where instead of polarity (positive or negative), modal
formulae are tested for monotonicity (upwards or downwards) in
propositional variables. The extension of \SQEMA\ based on the
stronger version of Ackermann's lemma is not presented in this
paper, but only briefly discussed in the concluding remarks in
Section \ref{ConcludingRemarks} and developed in more detail in
the sequel to the present paper \cite{CGVsqema3}.

\item it shows that the (suitably extended) modal language is rich
enough to enable the computation of the first-order equivalents of
many modal formulae.

\item it allows one to terminate the execution of the algorithm at any
stage and return a (hybrid) modal formula with partly eliminated
propositional variables, in an extension of the basic modal
language with inverse modalities and nominals.

\item it allows for a uniform proof of the correctness and canonicity
of the formulae on which the algorithm succeeds.
\end{itemize}

The rest of the paper is organized as follows: In Section 1 we
introduce the languages we will use, together with some basic
notions, and present a modal version of Ackermann's lemma. In
Section 2 we introduce the algorithm \SQEMA, followed in Section 3
by some examples of its execution on various input formulae. The
correctness of the algorithm and the canonicity of the formulae on
which it succeeds are proved in a uniform way in section 4, and
its completeness for the classes of Sahlqvist and inductive
formulae is shown in section 5. We conclude in section 6 by
discussing some extensions of \SQEMA.

\section{Preliminaries}

\subsection{Syntax, semantics and standard translations}

The \emph{basic modal language} $ML$ is built from $\bot$ and a countably infinite set of propositional
variables (or atomic propositions) $\textsf{PROP} = \{p_0, p_1, \ldots \}$ as usual:

$$ \phi ::= \bot \mid \top \mid p \mid \neg \phi \mid \phi \lor
\psi \mid \phi \land \psi \mid \dia \phi \mid \Box \phi.$$

An occurrence $\theta$ of a modal operator or a subformula in a formula $\phi$ has \emph{positive polarity} (or,
is \emph{positive}) if it is in the scope of an even number of negations; respectively, it has \emph{negative
polarity} (or, is \emph{negative}) if it is in the scope of an odd number of negations.

The connectives $\rightarrow$, and $\lra$ are defined as usual. Note that we take more logical connectives as
primitives than necessary, in order to avoid syntactic complications when introducing rules for each of them,
but whenever suitable, we will consider some of them definable in terms of the others, as usual.

A \emph{Kripke frame} is a pair $\mathfrak{F} = (W,R)$ where $W$ is a non-empty set of \emph{possible worlds} and $R
\subseteq W^{2}$ is an \emph{accessibility relation} between possible worlds. A \emph{Kripke model based on a
frame} $\mathfrak{F}$ is a pair $\mathcal{M} =(\mathfrak{F},V)$, where $V: \mathsf{PROP} \rightarrow 2^{W}$ is a
\emph{valuation} which assigns to every atomic proposition the set of possible worlds where it is true.

The \emph{truth of a formula} $\varphi$ \emph{at a possible world} $u$ \emph{of a Kripke model} $\mathcal{M}=
(W,R,V)$, denoted by $\mathcal{M},u \forces \varphi$, is defined recursively as follows:

$\mathcal{M},u  \forces \top$;

$\mathcal{M},u \forces p$ iff $u \in V(p)$;

$\mathcal{M}, u \forces \neg \phi$ iff $\mathcal{M}, u \not
\forces \phi$;

$\mathcal{M}, u \forces \phi \lor \psi$ iff $\mathcal{M}, u
\forces \phi$ or $\mathcal{M}, u \forces \psi$;

$\mathcal{M}, u \forces \phi \land \psi$ iff $\mathcal{M}, u
\forces \phi$ and $\mathcal{M}, u \forces \psi$;

$\mathcal{M}, u \forces \dia \varphi$ if $\mathcal{M},w \forces
\varphi$ for some $w \in W$ such that $Ruw$;

$\mathcal{M}, u \forces \Box \varphi$ if $\mathcal{M}, w \forces
\varphi$ for every $w \in W$ such that $Ruw$.\\
\\
A modal formula $\phi$ is:
\begin{itemize}
\item \emph{valid in a model} $\mathcal{M}$, denoted $\mathcal{M}
\forces \phi$, if it is true at every world of $\mathcal{M}$;

\item \emph{valid at a world $u$ in a frame} $\mathfrak{F}$, denoted
$\mathfrak{F}, u \forces \varphi$, if it is true at $u$ in every model
on $\mathfrak{F}$;

\item \emph{valid on a frame} $\mathfrak{F}$, denoted $\mathfrak{F}
\forces \varphi$, if it is valid in every model based on
$\mathfrak{F}$;

\item \emph{valid} if it is valid on every frame;

\item \emph{globally satisfiable on a frame $\mathfrak{F}$}, if there
exists a valuation $V$ such that $(\mathfrak{F}, V) \forces \phi$.
\end{itemize}

A \emph{general frame} is a structure $\mathfrak{F}=(W,R,\mathbb{W})$ where $(W,R)$ is a frame, and $\mathbb{W}$ is a
Boolean algebra of subsets of $2^W$, also closed under the modal operators. The elements of $\mathbb{W}$ are called
\emph{admissible sets} in $\mathfrak{F}$.

Given a general frame $\mathfrak{F}=( W,R,\mathbb{W})$, a \emph{model over} $\mathfrak{F}$ is a model over $(W,R)$ with the
valuation of the variables ranging over $\mathbb{W}$.

Given a modal formula $\psi$, a general frame $\mathfrak{F}$, and $w\in W$, we say that  $\psi$ is \textit{(locally)
valid at $w$ in} $\mathfrak{F}$, denoted $\mathfrak{F},w\forces \psi$, if $\psi $ is true at $w$ in every model over
$\mathfrak{F}$.

\medskip

Following \cite{vanBenthemBook} we define $L_{0}$ to be the first-order language with $=$, a binary predicate
$R$, and individual variables ${\sf VAR} =  \{x_{0},x_{1},\ldots \}$. Also, let $L_{1}$ be the extension of
$L_{0}$ with a set of unary predicates $\{P_{0},P_{1},\ldots \}$, corresponding to the atomic propositions
$\{p_{0},p_{1},\ldots \}$.

The formulae of $\mathrm{ML}$ are translated into $ L_{1}$ by means of the following \emph{standard translation}
function, $\mathrm{ST}(\cdot,\cdot)$, which takes as arguments an $ML$-formula together with a variable from
${\sf VAR}$:

$\mathrm{ST}(p_{i}, x):= P_{i}(x)$ for every $p_{i}\in {\sf PROP}$; $\mathrm{ST}(\neg \phi, x):= \neg
\mathrm{ST}(\phi, x)$;

$\mathrm{ST}(\phi \lor \psi, x):= \mathrm{ST}(\phi,x) \lor \mathrm{ST}(\psi,x)$; $\mathrm{ST}(\phi \land
\psi,x):= \mathrm{ST}(\phi,x) \land \mathrm{ST}(\psi,x)$;

$ST(\dia \phi, x) := \exists y (Rxy \land \mathrm{ST}(\phi, y))$, $ST(\Box \phi, x) := \forall y (Rxy
\rightarrow \mathrm{ST}(\phi, y))$,

where $y$ is the first variable in ${\sf VAR}$ not appearing in $ST(\phi,x)$.

\medskip

Now, for every  Kripke model $\mathcal{M}$, $w \in \mathcal{M}$ and $\phi \in \mathrm{ML}$:
\[
\mathcal{M}, w \forces \phi \ \mbox{ iff } \ \mathcal{M} \models
\mathrm{ST}(\phi, x)(x :=w),
\]
and
\[
\mathcal{M} \forces \phi \ \mbox{ iff } \ \mathcal{M} \models
\forall x \mathrm{ST}(\phi, x).
\]
Thus, on Kripke models the modal language is a fragment of $ L_{1}$.

Further, for every frame $\mathfrak{F}$, $w \in \mathfrak{F}$ and $\phi \in \mathrm{ML}$

\[\mathfrak{F}, w \forces \phi \ \mbox{ iff } \ \mathfrak{F} \models \forall
\overline{P} \mathrm{ST}(\phi, x)(x := w),
\]
and
\[
\mathfrak{F} \forces \phi \ \mbox{ iff } \ \mathfrak{F} \models \forall
\overline{P} \forall x \mathrm{ST}(\phi, x),
\]
where $\overline{P}$ is the tuple of all unary predicate symbols occurring in $\mathrm{ST}(\phi, x)$. Thus modal
formulae express universal monadic second-order conditions on frames.

\medskip

 A modal formula $\phi \in \mathrm{ML}$ is:
\begin{itemize}
\item \emph{locally first-order definable}, if there is a first-order
formula $\alpha(x)$ such that for every frame $\mathfrak{F}$ and $w
\in \mathfrak{F}$ it is the case that $\mathfrak{F}, w \forces \phi$ iff
$\mathfrak{F} \models \alpha(x:=w)$;

\item \emph{(globally) first-order definable}, if there is a first-order
sentence $\alpha$ such that for every frame $\mathfrak{F}$ it is the
case that $\mathfrak{F} \forces\phi$ iff $\mathfrak{F} \models\alpha$.
\end{itemize}

Two modal formulae are:
\begin{itemize}
\item
\emph{semantically equivalent} if they are true at exactly the
same states in the same models;
\item
\emph{locally frame equivalent} if they are valid at exactly the
same states in the same frames;
\item
\emph{frame equivalent} if they are valid on the same frames.
\end{itemize}

For the execution of the algorithm we enrich the language
$\mathrm{ML}$ by adding:
\begin{itemize}
\item the \emph{inverse modality} $\Box^{-1}$ with semantics
$\mathcal{M}, u \forces \Box^{-1}\phi$ \ iff \ $\mathcal{M}, w
\forces \phi$ for every $w \in \mathcal{M}$ such that $R^{-1}uw$,
i.e. $Rwu$. We extend the standard translation for the inverse
modality in the obvious way: $\mathrm{ST}(\Box^{-1} \phi, x ) :=
\forall y (Ryx \rightarrow \mathrm{ST}(\phi, y ))$, and define
$\dia^{-1}$ as the dual of $\Box^{-1}$.

\item \emph{nominals} (see \cite{GargovGoranko},\cite{MLBook}), a
special sort of propositional variables ${\sf Nom} = \{\nomi_{1},
\nomi_{2}, \ldots \}$ with admissible valuations restricted to
\emph{singletons}. The truth definition of nominals is: $(W,V), u
\forces \nomi$ iff $V(\nomi) = \{ u \}$. The standard translation:
$\mathrm{ST}(\nomi_{i}, x) := x = y_{i}$, where $y_0, y_1, \ldots$
are reserved variables associated with the nominals $\nomi_0,
\nomi_1, \ldots$.
\end{itemize}

The so extended modal language will be denoted by $\mathrm{ML}^+$.

Let $\mathcal{M}$ be a model and $\phi$ a formula from $\mathrm{ML}^+$. We denote $\linkerhaak \phi
\rechterhaak_{\mathcal{M}} = \{m \in \mathcal{M} : \mathcal{M},m \forces \phi \}$, i.e., $\linkerhaak \phi
\rechterhaak_{\mathcal{M}}$ is the \emph{extension} of the formula $\phi$ in the model $\mathcal{M}$.

A \emph{pure} formula in $\mathrm{ML}^+$ is a formula which contains no propositional variables, but which may
contain nominals. Every pure formula $\gamma$, is locally first-order definable by the formula $\forall
\overline{y} \mathrm{ST}(\gamma, x)$, where $\overline{y}$ is the tuple of all variables $y_i$ corresponding to
nominals $\nomi_i$ occurring in $\gamma$.

Let $A,B(p)$ be formulae in $\mathrm{ML}^+$. Then $B(A/p)$, hereafter also written as $B(A)$, is the formula
obtained from $B(p)$ by uniform substitution of $A$ for all occurrences of $p$.

Although formally $\mathrm{ML}^+$ is sufficient for the execution of our algorithm, it is sometimes useful to
further enrich the language with the \emph{universal modality} $[{\sf U}]$, with semantics $\mathcal{M}, u
\forces [{\sf U}] \phi$ iff $\mathcal{M},w \forces \phi$  for every $w \in \mathcal{M}$, i.e. iff $\mathcal{M}
\forces \phi$, and define $\langle {\sf U} \rangle$ as the dual of $[{\sf U}]$: $\mathcal{M}, u \forces \langle
{\sf U} \rangle \phi$ iff $\mathcal{M},w \forces \phi$ for some $w \in \mathcal{M}$ iff $\mathcal{M}, u \forces
\neg [{\sf U}] \neg \phi$. The standard translation is extended accordingly: $\mathrm{ST}([{\sf U}] \phi, x):=
\forall x \mathrm{ST}(\phi, x)$ and $\mathrm{ST}(\langle {\sf U} \rangle \phi, x):= \exists x \mathrm{ST}(\phi,
x)$. In particular, we use the universal modality in the next sub-section to re-phrase the modal version of
Ackermann's Lemma and compare it with the method of substitutions.

\subsection{Modal version of Ackermann's Lemma}

From now on we will work in $\mathrm{ML}^+$, unless otherwise specified. Let ${\sf AT} = {\sf PROP} \cup {\sf
NOM}$.

A formula $\phi$ is \emph{positive (negative) in a propositional variable $p$} if all occurrences of $p$ in
$\phi$ are in the scope of an even (odd) number of negations. $\phi$ is \emph{positive (negative)} if it is
positive (negative) in all propositional variables. Note that nominals are not taken in consideration here, and
a pure formula is both positive and negative.

Informally, a formula $\phi$ is \emph{upwards monotone} in a propositional variable $p$ if its truth is
preserved under extensions of the interpretation of $p$. Similarly, $\phi$ is said to be \emph{downwards
monotone} in $p$ if its truth is preserved under shrinkings of the interpretation of $p$. Here are the formal
definitions.

\begin{defi}{\bf (Monotonicity of a formula)}
A formula $\phi$ is {\it upwards monotone} in a propositional variable $p$ if for every frame $\mathfrak{F}$, all
states $w \in \mathfrak{F}$, and all valuations $V$ and $V'$ on $\mathfrak{F}$ such that $V(p) \subseteq V'(p)$ and
$V(q) = V'(q)$ for all $q \in {\sf AT}$, $q \neq p$, if $(\mathfrak{F},V), w \forces \phi$ then $(\mathfrak{F},V'), w
\forces \phi$, or, equivalently $\linkerhaak \phi \rechterhaak_{(\mathfrak{F},V)} \subseteq \linkerhaak \phi
\rechterhaak_{(\mathfrak{F},V')}$, where $\linkerhaak \phi \rechterhaak_{(\mathfrak{F},V)}$ denotes the extension of
$\phi$ in the Kripke model $(\mathfrak{F},V)$.

Similarly, a formula $\phi$ is {\it downwards monotone} in a propositional variable $p$ if for every frame
$\mathfrak{F}$, all states $w \in \mathfrak{F}$, and all valuations $V$ and $V'$ on $\mathfrak{F}$ such that $V'(p)
\subseteq V(p)$ and $V(q) = V'(q)$ for all $q \in {\sf AT}$, $q \neq p$, if $\mathfrak{F},V \forces \phi$ then
$\mathfrak{F},V' \forces \phi$, or, equivalently $\linkerhaak \phi \rechterhaak_{(\mathfrak{F},V)} \subseteq \linkerhaak
\phi \rechterhaak_{(\mathfrak{F},V')}$.

\end{defi}

Note that:

\begin{itemize}
\item the negation of a positive (resp. upwards
monotone) formula is a negative (resp. downwards monotone)
formula.

\item every formula positive in $p$ is upwards
monotone in $p$, and respectively, every formula negative in $p$
is downwards monotone in $p$.

\item the result of substitution of upwards (resp. downwards) monotone
formulae for the variables in any upwards monotone (in particular,
positive) formula is upwards (resp. downwards) monotone. In
particular, $\phi(p)$ is upwards (resp. downwards) monotone in $p$
iff $\phi(\lnot p)$ is downwards (resp. upwards) monotone in $p$.

\end{itemize}

\begin{lem}[Modal Ackermann Lemma]
Let $A,B(p)$ be formulae in $\mathrm{ML}^+$ such that the propositional variable $p$ does not occur in $A$ and
$B(p)$ is negative in $p$. Then for any model $\mathcal{M}$, $\mathcal{M} \forces B(A)$ iff $\mathcal{M}'
\forces (A \ra p) \land B(p)$ for some model $\mathcal{M}'$ which may only differ from $\mathcal{M}$ on the
valuation of $p$.\label{ModalAckermann}
\end{lem}

\proof If $\mathcal{M} \forces B(A)$, then $\mathcal{M}' \forces (A \ra p) \land B(p)$ for a model
$\mathcal{M}'$ such that $\linkerhaak p \rechterhaak_{\mathcal{M}'} = \linkerhaak A \rechterhaak_{\mathcal{M}}$.
Conversely, if $\mathcal{M}' \forces (A \ra p) \land B(p)$ for some  model $\mathcal{M}'$ then $\mathcal{M}'
\forces B(A/p)$ since  $B(p)$ is downwards monotone. Therefore, $\mathcal{M} \forces B(A/p)$.\qed

An analogue of this lemma can be formulated for positive formulae $B$, too. We note that a somewhat different
version of Ackermann's lemma has been proved for modal first-order formulae in \cite{Szalas4}, where it is also
applied to some modal formulae.

\subsection{Ackermann's Lemma and the substitution method}

Note that the contrapositive form of the modal Ackermann's lemma (after replacing $\lnot B$ with $B$) is
equivalent to:
\[
\forall p([{\sf U}](A\rightarrow p)\rightarrow B(p))\ \equiv \
B(A/p),
\]
for any modal formula $A$ not containing $p$, and a modal formula $B$ which is positive with respect to $p$. To
see that equivalence, note that $[{\sf U}](A\rightarrow p)$ is true in a model $\mathcal{M}$ iff $\linkerhaak A
\rechterhaak_{\mathcal{M}} \subseteq \linkerhaak p \rechterhaak_{\mathcal{M}}$.

The statement above can be interpreted as follows: $[{\sf U}](A\rightarrow p)\rightarrow B(p)$ is valid in a
given frame iff $B(p)$ is true for the `minimal' valuation satisfying the antecedent, viz. $A$. \emph{This is
precisely the technical idea at the heart of the substitution method!}

\section{The algorithm SQEMA}
In this section we formally introduce the core algorithm \SQEMA.
To reduce technical overhead in the exposition, the algorithm will
be presented in the basic modal language. In \cite{CGVsqema2}
\SQEMA\ is extended to arbitrary polyadic and hybrid multi-modal
languages.

\subsection{The core algorithm}

Given a modal formula $\phi$ as input, we transform $\neg \phi$ into negation normal form. We then apply the
algorithm's transformation rules (all of which preserve local frame-equivalence), to formulae of the form
$\alpha \ra \beta$, with $\alpha$ and $\beta$ both in negation normal form, which we call
`equations'\footnote{Because the algorithm resembles a procedure of solving systems of linear algebraic
equations by Gaussian elimination.}. All equations are interpreted as global statements (i.e., valid in the
whole model). The algorithm works with systems of equations. Each transformation rule transforms an equation
into one or more new equations. The Ackermann-rule, based on Ackermann's lemma, is applicable to a whole system
of equations. The idea is to choose a propositional variable and transform the current system of equations,
using the transformation rules presented further, into a system to which the Ackermann-rule can be applied with
respect to the chosen variable, thus eliminating it. This process is then repeated until all propositional
variables are eliminated. As will be seen (see example \ref{ExOrd}), the success of the algorithm may in general
depend upon the order in which this elimination of propositional variables is attempted. The algorithm
accordingly makes provision for trying different orders of elimination.

Now for a more formal description. The algorithm takes as input a modal formula $\phi$ and proceeds as follows:

\begin{description}

\item[Step 1] Negate $\phi$ and rewrite $\neg \phi$ in negation
normal form by eliminating the connectives `$\ra$' and `$\lra$',
and by driving all negation signs inwards until they appear only
directly in front of propositional variables.

Then distribute diamonds and conjunctions over disjunctions as
much as possible, using the equivalences $\dia(\phi \lor \psi)
\equiv (\dia\phi \lor \dia\psi)$ and $(\varphi \vee \psi )\wedge
\theta \equiv (\varphi \wedge \theta )\vee (\psi \wedge \theta )$,
in order to obtain $\bigvee \alpha_k$, where no further
distribution of diamonds and conjunctions over disjunctions is
possible in any $\alpha_k$.

The algorithm now proceeds on each of the disjuncts, $\alpha_k$
separately, as follows:

\item[Step 2] Rewrite $\alpha_k$ as $\nomi \ra \alpha_k$, where
$\nomi$ is a fixed, reserved nominal, not occurring in any
$\alpha_k$, and only used to denote the initial current state.
This is the only equation in the initial system.

\item[Step 3] Eliminate every propositional variable in which
the system is positive or negative, by replacing it with $\top$ or $\bot$, respectively.

\item[Step 4]
If there are propositional variables remaining in equations of the
system, choose one to eliminate, say $p$, the elimination of which
has not been attempted yet. Proceed to step 5. (So far this choice
is made non-deterministically; some heuristics will be suggested
in a sequel paper.) If all remaining variables have been attempted
and Step 5 has failed, backtrack and attempt another order of
elimination.

If all orders of elimination and all remaining variables have been attempted and step 5 has failed, report
failure.

If all propositional variables have been eliminated from the system, proceed to Step 6.

\item[Step 5] The goal now is, by applying the transformation rules listed below, to rewrite the system of equations so that
the Ackermann-rule becomes applicable with respect to the chosen variable $p$ in order to eliminate it. Thus,
the current goal is to transform the system into one in which every equation is either negative in $p$, or of
the form $\alpha \ra p$, with $p$ not occurring in $\alpha$, i.e. to `extract' $p$ and `solve' for it.

If this fails, backtrack, change the polarity of $p$ by substituting $\neg p$ for it everywhere, and attempt
again to prepare for the Ackermann-rule.

If this fails again, or after the completion of this step, return to Step 4.

\item[Step 6] If this step is reached by all the branches of the execution it means that all propositional variables have been
successfully eliminated from all systems resulting from the input formula. What remains now is to return the
desired first-order equivalent. In each system, take the conjunction of all equations to obtain a formula ${\sf
pure}$, and form the formula $\forall \overline{y} \exists x_0 {\rm ST}(\neg {\sf pure}, x_0)$, where
$\overline{y}$ is the tuple of all occurring variables corresponding to nominals, but with $y_i$ (corresponding
to the designated current state nominal $\nomi$) left free if the local correspondent is to be computed. Then
take the conjunction of these translations over the systems on all disjunctive branches. For motivation of the
correctness of this translation the reader is referred to the examples in the following section as well as the
correctness proof in section \ref{CorrectnessSection}.

Return the result, which is the (local) first-order condition corresponding to the input formula.

\end{description}

Note that, if a right order of the elimination of the variables is guessed (i.e. chosen non-deterministically)
rather than all possible orders explored sequentially, this algorithm runs in nondeterministic polynomial time.
However, we believe that additional rules which determine the right order of elimination (if any) can reduce the
complexity to PTIME.

\subsection{Transformation Rules}

The transformation rules used by \SQEMA\ are listed below. Note
that these are \emph{rewriting rules}, i.e. they replace the
equations to which they apply.\footnote{Rajeev Gore has remarked
that these transformation rules are reminiscent of the rules in
display logics.}

\medskip

\textbf{I. Rules for the logical connectives:}

\begin{center}

\begin{tabular}{c}
$\wedge $-rule:\hspace{1.5cm}$
\begin{array}{c}
\beta \rightarrow \gamma \wedge \delta  \\
\Downarrow  \\
\beta \rightarrow \gamma ,\beta \rightarrow \delta
\\
\end{array}
 $
\\~\\
\end{tabular}

\begin{tabular}{cc}
Left-shift $\vee $-rule:

$\begin{array}{c}
\beta \rightarrow \gamma \vee \delta  \\
\Downarrow  \\
(\beta \wedge \lnot \gamma )\rightarrow \delta
\\
\end{array}
\hspace{1cm}
 $ & Right-shift $\vee $-rule: $
\begin{array}{c}
(\beta \wedge \lnot \gamma )\rightarrow \delta  \\
\Downarrow  \\
\beta \rightarrow \gamma \vee \delta
\\~\\
\end{array}
$
\\~\\
\end{tabular}
\\~\\

\begin{tabular}{cc}
Left-shift  $\Box $-rule: $
\begin{array}{c}
\gamma \rightarrow \Box \delta  \\
\Downarrow  \\
\Diamond ^{-1}\gamma \rightarrow \delta
\end{array}
\hspace{1cm}
 $ & Right-shift $\Box $-rule: $
\begin{array}{c}
\Diamond ^{-1}\gamma \rightarrow \delta  \\
\Downarrow  \\
\gamma \rightarrow \Box \delta
\\
\end{array}
$
\\~\\
\end{tabular}
\\~\\

\begin{tabular}{cc}
$\Diamond $-rule: & $
\begin{array}{c}
\nomj\rightarrow \Diamond \gamma  \\
\Downarrow  \\
\nomj \rightarrow \Diamond \nomk, \ \nomk \rightarrow \gamma \\
\mbox{where $\nomj$ is any nominal and $\nomk$ is a new nominal.}
\\
\end{array}
$
\end{tabular}
\end{center}

Sometimes we will write $R \nomj \nomk$ as an abbreviation for $\nomj \rightarrow \Diamond \nomk$.

\medskip

\textbf{II. Ackermann-Rule:}  This rule is based on the equivalence given in Ackermann's Lemma. It works, not on
a single equation, but by transforming a whole set of equations as follows:

\begin{center}
\begin{tabular}{ccc}
$\left \| \begin{array}[l]{l}

\alpha_{1}\rightarrow p,
\\ \ldots
\\ \alpha_{n}\rightarrow p,
\\
\\ \beta_{1}(p),
\\ \ldots
\\ \beta_{m}(p),
\end{array}
\right.$ &

$\ \ \ \Rightarrow \ \ \ $

& $\left \| \begin{array}[l]{l}

\beta_{1}[(\alpha _{1}\vee \ldots \vee \alpha_{n})/p],
\\ \ldots
\\ \beta_{m}[(\alpha _{1}\vee \ldots \vee \alpha_{n})/p].
\end{array}
\right.$
\end{tabular}
\end{center}

where:

\begin{enumerate}
\item $p$ does not occur in $\alpha_{1},\ldots,\alpha_{n}$;
\item each of $\beta_{1},\ldots,\beta_{m}$ is negative in $p$;
\item no other equations in the system contain $p$.
\end{enumerate}

Hereafter, we will refer to the formulae $\alpha_{1},\ldots,\alpha_{n}$ in an application of the Ackermann-rule
as \emph{$\alpha$-formulae}, and to $\beta_{1},\ldots,\beta_{m}$ as \emph{$\beta$-formulae}.

\medskip

\textbf{III. Polarity switching rule:} Switch the polarity of every occurrence of a chosen variable $p$ within
the current system, i.e. replace $\neg p$ by $p$ and $p$ by $\neg p$ for every occurrence of $p$ not prefixed by
$\neg$.

\medskip

\textbf{IV. Auxiliary Rules:} These rules are intended to provide the algorithm with some propositional
reasoning capabilities and to effect the duality between the modal operators.

\begin{enumerate}
\item  Commutativity and associativity of $\land$ and $\lor$ (tacitly used).

\item  Replace $\gamma \vee \lnot \gamma $ with $\top ,$ and $\gamma \wedge
\lnot \gamma $ with $\bot $.

\item  Replace $\gamma \vee \top $ with $\top ,$ and $\gamma \vee \bot $
with $\gamma $.

\item  Replace $\gamma \wedge \top $ with $\gamma ,$ and $\gamma \wedge \bot
$ with $\bot $.

\item  Replace $\gamma \rightarrow \bot $ with $\lnot \gamma $ and $\gamma
\rightarrow \top $ with $\top $.

\item  Replace $\bot \rightarrow \gamma $ with $\top $ and $\top \rightarrow
\gamma $ with $\gamma $.

\item  Replace $\neg \Diamond \neg $ with $\Box $ and $\neg \Box \neg $ with $\Diamond$.

\end{enumerate}

Note that, apart from the polarity switching rule, no transformation rule changes the polarity of any occurrence
of a propositional variable.

\section{Some examples of the execution of \SQEMA} \label{ExampleSection}

In this section we illustrate the execution of \SQEMA\ on several formulae and discuss some of its features.

\begin{exa}
Consider the formula $\dia \Box p \ra \Box \dia p$.
\begin{description}
\item[Step 1] Negating, we obtain $ \dia \Box p \land \dia \Box \neg p $.

\item[Step 2] The initial system of equations:
$$\left \| \nomi \ra (\dia \Box p \land \dia \Box \neg p) \right..$$

\item[Step 3] The formula is neither positive nor negative
in $p$, the only occurring propositional variable.

\item[Step 4] We choose $p$ to eliminate --- our only option.

\item[Step 5] We will try to transform the system, using the
rules, so that the Ackermann-rule becomes applicable.

Applying the $\land$-rule gives

$$\left \| \begin{array}[l]{l}
\nomi \ra \dia \Box p\\
\nomi \ra \dia \Box \neg p
\end{array}
\right.,$$

Applying first the $\dia$-rule and then the $\Box$-rule to the
first equation yields:

\begin{center}
\begin{tabular}{ccc}
$\left \| \begin{array}[l]{l}
R \nomi \nomj\\
\nomj \ra \Box p\\
\nomi \ra \dia \Box \neg p
\end{array}
\right.$,
\end{tabular}
\end{center}

and then

\begin{center}
\begin{tabular}{ccc}
$\left \| \begin{array}[l]{l}
R \nomi \nomj\\
\dia^{-1} \nomj \ra  p\\
\nomi \ra \dia \Box \neg p
\end{array}
\right.$.
\end{tabular}
\end{center}

The  Ackermann-rule is now applicable, yielding the system
$$\left \| \begin{array}[l]{l}
R \nomi \nomj\\
\nomi \ra \dia \Box \neg (\dia^{-1} \nomj)
\end{array}
\right.$$

All propositional variables have been successfully eliminated, so
we  proceed to step 6.

\item[Step 6] Taking the conjunction of the equations gives
$$R \nomi \nomj \land (\nomi \ra \dia \Box \neg (\dia^{-1}\nomj)).$$

Negating we obtain
$$R \nomi \nomj \ra  (\nomi \land \Box \dia \dia^{-1} \nomj),$$

which, translated, becomes

$$ \forall y_j \exists x_0 [Ry_i y_j \ra (x_0 = y_i) \land \forall
y (Rx_{0}y \ra \exists u (Ryu \land \exists v (Rvu \land v =
y_j)))],$$

and simplifies to

$$ \forall y_j [Ry_i y_j \ra \forall y (Ry_i y \ra \exists u (Ryu
\land Ry_ju ))] $$

defining the Church-Rosser property, as expected. Note that the
variable $y_i$ occurs free and corresponds to the nominal
$\mathbf{i}$, which we interpret as the current state. Hence we
obtain a \emph{local} property. Also, we directly translate the
abbreviation $R \nomi \nomj$ as $Ry_i y_j$, since ${\rm ST}(\nomi
\ra \dia \nomj, x_0)$ is $x_0 = y_i \ra \exists z (Rx_0 z \land z
= y_j)$, which clearly simplifies.
\end{description}
\end{exa}

\begin{exa}
Consider the formula $p \land \Box(\dia p \ra \Box q) \ra \dia
\Box \Box q$. Note that it is \emph{not equivalent to a Sahlqvist
formula} (see \cite{GorankoVakarelov3}).

\begin{description}
\item[Step 1] Negating we obtain $p \land \Box(\Box \neg p \lor
\Box q) \land \Box \dia \dia \neg q$.

\item[Step 2] $\nomi \ra [p \land \Box(\Box \neg p \lor \Box q)
\land \Box \dia \dia \neg q]$.

\item[Step 3] Nothing to do, as the system is neither positive nor negative
in $p$ or $q$.

\item[Step 4] Choose $p$ to eliminate.

\item[Step 5] Applying the $\land$-rule twice, we get

\begin{center}
\begin{tabular}{ccc}
$\left \| \begin{array}[l]{l}
\nomi \ra p\\
\nomi \ra \Box(\Box \neg p \lor \Box q))\\
\nomi \ra \Box \dia \dia \neg q
\end{array} \right..$
\end{tabular}
\end{center}

The system is now ready for the application of the Ackermann-rule, as $p$ has been successfully isolated, and
$\nomi \ra \Box(\Box \neg p \lor \Box q))$ is negative in $p$:

$$\left \|
\begin{array}[l]{l}
\nomi \ra \Box(\Box \neg \nomi \lor \Box q))\\
\nomi \ra \Box \dia \dia \neg q
\end{array} \right..$$

\item[Step 4] Choose $q$, the only remaining option, to eliminate
next.

\item[Step 5] Successively applying the $\Box$-rule, Left-shift
$\lor$-rule and $\Box$-rule again, we get
$$\left \| \begin{array}[l]{l}
\dia^{-1}(\dia^{-1} \nomi \land \dia \nomi) \ra  q\\
\nomi \ra \Box \dia \dia \neg q
\end{array} \right..$$

Applying the Ackermann-rule to eliminate $q$, and rewriting in
negation normal form yields:
$$\left \| \begin{array}[l]{l}
\nomi \ra \Box \dia \dia [\Box^{-1}(\Box^{-1} \neg \nomi \lor \Box
\neg \nomi)]
\end{array} \right..
$$

\item[Step 6] $(\nomi \ra \Box \dia \dia [\Box^{-1}(\Box^{-1} \neg
\nomi \lor \Box \neg \nomi))$, when negated becomes $\nomi \land
\dia \Box \Box [\dia^{-1}(\dia^{-1} \nomi \land \dia \nomi).$
Translating into first-order logic we obtain

 $$\begin{array}[l]{l}
 \exists x_0 [x_0 = y_i \land \exists z_1(Rx_{0}z_{1} \land \forall z_2(Rz_1z_2 \ra
 \forall z_3 (Rz_2z_3 \ra \exists u_1[Ru_1z_3\\
 \land \exists u_2 (Ru_2u_1 \land u_2 = y_i) \land
 \exists u_3 (Ru_1u_3 \land u_3 = y_i )])))]. \end{array}$$

\end{description}
Note that, in this example, the order of elimination of the propositional variables is inessential, as
eliminating $q$ first and then $p$ works equally well.
\end{exa}

\begin{exa} \label{ExOrd}
Consider the formula $\Box(\Box p \leftrightarrow q) \ra p$. The current implementations of both SCAN
(\texttt{http://www.mpi-inf.mpg.de/\~{}ohlbach/scan/corr\_form.html}) and DLS
(\texttt{http://www.ida.liu.se/labs/kplab/projects/dls}) fail on it.  Let's see what \SQEMA\ does with it:

\begin{description}
\item[Step 1] $\Box((\dia \neg p \lor q) \land (\neg q \lor \Box p)) \land \neg p$.

\item[Step 2]
$$\left \| \begin{array}[l]{l}
\nomi \ra \Box((\dia \neg p \lor q) \land (\neg q \lor \Box p)) \land \neg p
\end{array} \right..
$$

\item[Step 3] All variables occur both positively and negatively, so we go on to step 4.

\item[Step 4] Choose $q$ to eliminate.

\item[Step 5] Applying the $\land$-rule and the $\Box$-rule we transform the system into
$$\left \| \begin{array}[l]{l}
\dia^{-1} \nomi \ra (\dia \neg p \lor q)\\
\dia^{-1} \nomi \ra (\neg q \lor \Box p)\\
\nomi \ra \neg p
\end{array} \right..
$$

Applying the Left Shift $\lor$-rule to the first equation yields
$$\left \| \begin{array}[l]{l}
(\dia^{-1} \nomi \land \Box p) \ra  q\\
\dia^{-1} \nomi \ra (\neg q \lor \Box p)\\
\nomi \ra \neg p
\end{array} \right..
$$

to which the Ackermann-rule is applicable with respect to $q$. This gives
$$\left \| \begin{array}[l]{l}
\dia^{-1} \nomi \ra (\neg \dia^{-1} \nomi \lor \neg \Box p \lor \Box p)\\
\nomi \ra \neg p
\end{array} \right..
$$

The first equation in the system is now a tautology and may be removed (we interpret systems of equations
conjunctively), yielding the system
$$\left \| \begin{array}[l]{l}
\nomi \ra \neg p
\end{array} \right.,
$$

in which $p$ may be replaced by $\bot$ since it occurs only negatively, resulting in the system
$$\left \| \begin{array}[l]{l}
\top
\end{array} \right.
$$

\item[Step 6] Negating we obtain $\bot$.
\end{description}

Some remarks are in order here. Firstly, note that the success of the algorithm may depend essentially upon the
ability to do some propositional reasoning. Particularly, it had to recognize $\dia^{-1} \nomi \ra (\neg
\dia^{-1} \nomi \lor \neg \Box p \lor \Box p)$ as a tautology. If this had been written in a slightly different
form, however, say as $\dia^{-1} \nomi \ra (\neg \dia^{-1} \nomi \lor \dia \neg p \lor \Box p)$, its
tautological status may not be recognized so easily. This is where a stronger version of Ackermann-rule (see
Section \ref{ConcludingRemarks}), which involves testing for \emph{monotonicity}, rather than polarity would
prove to be useful. For, testing the consequent for (upward) monotonicity would give the answer `yes' --- this
is, post hoc, easy to see, since we already know that the consequent is a tautology. This would allow us, after
changing the polarity of $p$, to apply the stronger, monotonicity based, version of the Ackermann-rule to the
system
$$\left \| \begin{array}[l]{l}
\dia^{-1} \nomi \ra (\neg \dia^{-1} \nomi \lor \dia p \lor \Box \neg p)\\
\nomi \ra p
\end{array} \right.,
$$

and obtain
$$\left \| \begin{array}[l]{l}
\dia^{-1} \nomi \ra (\Box^{-1} \neg \nomi \lor \dia  \nomi \lor \Box \neg \nomi)\\
\end{array} \right.,
$$

which simplifies to
$$\left \| \begin{array}[l]{l}
\dia^{-1} \nomi \ra (\Box^{-1} \neg \nomi \lor \dia \nomi \lor \neg \dia \nomi)\\
\end{array} \right.,
$$

which, as before, is a tautology, yielding, in  step 6, the first order equivalent $\bot$.

Secondly, suppose that we had tried to eliminate $p$ first. Note that we will gain nothing by changing the
polarity of $p$ for we cannot get the occurrence of $p$ `out' under the diamond in $\dia^{-1} \nomi \ra (\dia p
\lor q)$. Indeed, the system may be transformed to become

$$\left \| \begin{array}[l]{l}
\dia^{-1} \nomi \ra (\dia \neg p \lor q)\\
\dia^{-1}(\dia^{-1} \nomi \land q) \ra  p\\
\nomi \ra \neg p
\end{array} \right.,
$$

to which we may apply the  Ackermann-rule with respect to $p$ and obtain

$$\left \| \begin{array}[l]{l}
\dia^{-1} \nomi \ra (\dia \neg (\dia^{-1}(\dia^{-1} \nomi \land q)) \lor q)\\
\nomi \ra \neg (\dia^{-1}(\dia^{-1} \nomi \land q))
\end{array} \right..
$$

It is now not difficult to see that \SQEMA\ gets stuck on this system.

Moral: the order of elimination does matter sometimes, that is why the algorithm incorporates the backtrack
option. Of course, theoretically this may lead to an exponential increase of the number of steps, but in
practice this apparently does not happen if suitable heuristics or additional rules can be applied to determine
the right order of elimination.
\end{exa}

\section{Correctness and canonicity of SQEMA} \label{CorrectnessSection}

\subsection{Modal formulae as operators on general frames}

For what follows, it is useful to think of modal formulae as set-theoretic operators.  If $\psi \in ML^{+}$, let
${\sf PROP}(\psi )=\{q_{1},\ldots ,q_{m}\} $ be the propositional variables occurring in $\psi$, ${\sf
NOM}(\psi) = \{\mathbf{i}_{1},\ldots ,\mathbf{i}_{n} \}$ the nominals occurring in $\psi$, and ${\sf AT}(\psi) =
{\sf PROP}(\psi) \cup {\sf NOM}(\psi)$. Recall that with every model $\mathcal{M}=\left\langle
W,R,V\right\rangle$ and a modal formula $\psi $ in $\mathrm{ML}$ we associate the set $[\![\psi
]\!]_{\mathcal{M}}$ denoting the extension of the formula $\psi$ in the model $\mathcal{M}$. Clearly, $[\![\psi
]\!]_{\mathcal{M}}$ only depends on the valuation of ${\sf PROP}(\psi)$ in $\mathcal{M}$, and therefore
$[\![\psi ]\!]$ defines an operator from $\mathcal{P}(W)^{m}$ to $ \mathcal{P}(W).$ Likewise for formulae of the
extended language $ML^{+}$, with the amendment that if ${\sf PROP}(\psi) = \{q_{1},\ldots ,q_{m} \}$ and ${\sf
NOM}(\psi) = \{\mathbf{i}_{1},\ldots ,\mathbf{i}_{n} \}$, then $[\![\psi ]\!]:\mathcal{P} (W)^{m}\times
W^{n}\rightarrow \mathcal{P}(W)$. Whenever appropriate, we will simply identify formulae with the operators they
define. For instance, if $\phi = \Box^{-1} p \lor \dia \mathbf{j}$, $x \in W$ and $A \subseteq W$, we will write
$\Box^{-1} A \lor \dia \{x \}$ or $\Box^{-1} A \cup \dia \{x \}$ to denote the extension of $\phi$ when the
valuation of $p$ is $A$ and that of $\mathbf{j}$ is $\{ x \}$. If we have to specify the relation, we will write
$\langle R \rangle$ and $\lbrak R \rbrak$ for $\dia A$ and $\Box A$ respectively. Also $R(A)$ (respectively
$R^{-1}(A)$) will be used to denote the set of successors (respectively predecessors) of points in $A$. We will
be sloppy and write $R(x)$ (respectively $R^{-1}(x)$) for $R(\{ x \})$ (respectively $R^{-1}(\{ x \})$).

We will now extend (\emph{ad hoc}) the notion of local validity at a state $w$ of a general frame
$\mathfrak{F}=\langle W,R,\mathbb{W}\rangle$ for the basic language $\mathrm{ML}$, to formulae of the extended
language ${ML}^{+}$ as follows: For a formula $\phi \in ML^{+}$, it is the case that ${\mathfrak{F},w \Vdash\psi}$
if $ \mathcal{M}{,w\Vdash\psi}$ for every model $\mathcal{M\in }M(\mathfrak{F})$ extended in a standard way to the
inverse modalities, and where all nominals can take as values any $v \in W$, \emph{except for the special
nominal} $\mathbf{i}$ \emph{\ which is only interpreted in the current state} $w$. Note that, in doing so, we
lose the property that the extension of any formula in a model based on a general frame will itself be an
admissible set (i.e. a member of $\mathbb{W}$).

\subsection{Topology on Descriptive Frames}

Hereafter, unless otherwise specified, we only deal with general (and in particular, descriptive) frames in
$\mathrm{ML}$. For the basic topological concepts used in this section, the reader may consult any standard
reference on topology, e.g. \cite{Willard}.

With every general frame $\mathfrak{F}=\left\langle W,R,\mathbb{W}\right\rangle$, we associate a topological space
$(W, T(\mathfrak{F}))$, where $\mathbb{W}$ is taken as a base of clopen sets for the topology $T(\mathfrak{F})$. Let
$\mathbf{C}(\mathbb{W})$ denote the set of sets closed with respect to $T(\mathfrak{F})$.

A general frame $\mathfrak{F}=\left\langle W,R, \mathbb{W}\right\rangle$ is said to be \emph{differentiated} if for
every $x,y \in W$, $x \neq y$, there exists $X \in \mathbb{W}$ such that $x \in X$ and $y \not \in X$
(equivalently, if $T({\mathfrak{F}})$ is Hausdorff); \emph{tight} if for all $x, y \in W$ it is the case that $Rxy$
iff $x \in \bigcap \{\langle R \rangle (Y) : Y \in \mathbb{W} \textrm{ and } y \in Y \}$ (equivalently, if $R$
is point-closed, i.e. R(\{x\}) is closed for every $x \in W$); \emph{compact} if every family of admissible sets
from $\mathbb{W}$ with the finite intersection property (FIP) has non empty intersection (equivalently, if
$T(\mathfrak{F})$ is compact). $\mathfrak{F}$ is called \emph{descriptive} it is differentiated, tight and compact.

\begin{defi}
A formula $\psi \in \mathrm{ML}$ is \emph{locally d-persistent} if for every descriptive frame
$\mathfrak{F}=\left\langle W,R,\mathbb{W}\right\rangle $ {and }$w\in W$,
\begin{equation*}
{\mathfrak{F},w\Vdash\psi \ }\mbox{{iff}}{\ }F{,w\Vdash\psi,}
\end{equation*}
where $F={\left\langle W,R\right\rangle .}$
\end{defi}

Note that we only talk about local d-persistence of formulae from the basic language $\mathrm{ML}$. It is well
known (see e.g. \cite{MLBook}) that (local) d-persistence implies canonicity of formulae in $\mathrm{ML}$
because the canonical general frame for every normal modal logic is descriptive, and hence every d-persistent
axiom is valid in its canonical frame.

\begin{defi}
A formula $\gamma =\gamma (p_{1},\ldots ,p_{n})$ from $ML^{+}$ is a \emph{closed operator in $\mathrm{ML}$}, if
for every descriptive frame $ \mathfrak{F}=\left\langle W,R,\mathbb{W} \right\rangle $, if $P_{1},\ldots ,P_{n}\in
\mathbf{C(}\mathbb{W)}$ and any singletons are assigned to the nominals in $\gamma$, then $\gamma (P_{1},\ldots
,P_{n})\in \mathbf{C(} \mathbb{W)}$, i.e. when applied to closed sets in a descriptive frame it produces a
closed set; $\gamma$ is a \emph{closed formula in $\mathrm{ML}$} if whenever applied to \emph{admissible} sets
in any descriptive frame it produces a closed set. Thus, if a formula is a closed operator, then it is a closed
formula, but not necessarily vice versa.

Similarly, a formula from $ML^{+}$ is an \emph{open operator in $\mathrm{ML}$} if whenever applied to open sets
in a descriptive frame it produces an open set; it is an \emph{open formula in $\mathrm{ML}$} if whenever
applied to admissible sets it produces an open set.
\end{defi}

Note that the operators $\dia$ and $\dia^{-1}$ distribute over arbitrary unions, and $\Box$ and $\Box^{-1}$
distribute over arbitrary intersections. Since every closed set can be obtained as the intersection admissible
sets and each open set as the union of admissible sets,  and  $\dia$ and $\Box$ applied to admissible sets yield
admissible sets, it follows that $\dia p$ is an open operator and $\Box p$ is a closed operator, and this holds
even for arbitrary general frames.

The proof of the next lemma, originally from \cite{Esakia}, can also be found in \cite{SambinVacarro2} or
\cite{MLBook}.

\begin{lem}[Esakia's Lemma for $\dia$]
Let $\mathfrak{F}$ be a descriptive frame. Then for any downward directed family of nonempty closed sets $\{ C_i : i
\in I \}$ from $T(\mathfrak{F})$, it is the case that $\dia \bigcap_{i \in I} C_i = \bigcap_{i \in I} \dia
C_i$.\label{DiamondEsakiaLemma}
\end{lem}

\begin{cor}
$\dia p$ is a closed operator in $ML$. \label{diaClosedCor}
\end{cor}

\begin{lem}[Esakia's Lemma for $\dia^{-1}$ in $ML$]
Let $\mathfrak{F}$ be a descriptive frame. Then  $\dia^{-1} \bigcap_{i \in I} C_i = \bigcap_{i \in I} \dia^{-1} C_i$
for any downwards directed family of nonempty closed sets $\{ C_i : i \in I \}$ from
$T(\mathfrak{F})$.\label{InvDiaEsakiaLemma}
\end{lem}

\proof The inclusion $\dia^{-1} \bigcap_{i \in I} C_i \subseteq \bigcap_{i \in I} \dia^{-1} C_i$ is trivial, so
suppose that $x_0 \not \in \dia^{-1} \bigcap_{i \in I} C_i$, i.e. $\langle R \rangle (x_0) \cap \bigcap_{i \in
I} C_i = \emptyset$. Now $ \langle R \rangle (x_0)$ is closed by corollary \ref{diaClosedCor} and the fact that
singletons are closed in descriptive frames. Hence $\{ \langle R \rangle (x_0)\} \cup \{ C_i : i \in I \}$ is a
family of closed subsets with empty intersection which, by compactness, cannot have the FIP. Thus there is a
finite subfamily $\{C_1, \ldots, C_n \} \subseteq \{ C_i : i \in I \}$ such that $\langle R \rangle (x_0) \cap
C_1 \cap \cdots \cap C_n = \emptyset$. Since $\{ C_i : i \in I \}$ is downwards directed, it follows that there
that there exists a $C \in \{ C_i : i \in I \}$ such that $C \subseteq \bigcap \{C_1, \ldots, C_n \}$ and
$\langle R \rangle (x_0) \cap C = \emptyset$. But then $x_0 \not \in \dia^{-1}C$, and hence $x_0 \not \in
\bigcap_{i \in I} \dia^{-1}C_i$.\qed

\begin{lem}
$\dia^{-1}p$ is a closed operator in $ML$.
\label{InvDiaClosedOperatorLemma}
\end{lem}

\proof The proof is adapted from \cite{GorankoVakarelov3}. Let $\mathfrak{F}=\left\langle W,R,
\mathbb{W}\right\rangle$ be a descriptive frame in $ML$. We have to show that for any closed $A \subseteq W$ it
is the case that $\dia^{-1}(A) = \bigcap \{ B \in \mathbb{W}: \dia^{-1}A \subseteq B\}$. Note that $\dia^{-1}(A)
= R(A)$. The inclusion from left to right is trivial. In order to prove the right-to-left inclusion, suppose
that $x_0 \not \in \dia^{-1}(A)$, i.e. for all $y \in A$ it is not the case that $Ryx_0$. By the
point-closedness of $R$ we have that $R(y) = \bigcap\{B \in \mathbb{W} : y \in \Box B \}$. Then for each $y \in
A$ there must exist a $B^{y} \in \mathbb{W}$ such that $y \in \Box B^{y}$ and $x_0 \not \in B^{y}$, and hence $A
\subseteq \bigcup \{\Box B^{y} : y \in A\}$. Therefore $\{\Box B^{y} : y \in A\}$ is an open cover of the closed
set $A$, so by compactness there exists a finite subcover $\Box B_1, \ldots, \Box B_n$. Then $A \subseteq \Box
B_1 \cup \cdots \cup \Box B_n$ and $x_0 \not \in B_i, 1 \leq i \leq n$. Since $\dia^{-1}$ distributes over
arbitrary unions, we then have $\dia^{-1}A \subseteq \dia^{-1}\Box B_1 \cup \cdots \cup \dia^{-1}\Box B_n$. And,
since for any $X \subseteq W$, $\dia^{-1} \Box X \subseteq X$, we have $\dia^{-1}A \subseteq B_1 \cup \cdots
\cup B_n$. So we have found an admissible set containing $\dia^{-1}A$, not containing $x_0$, and hence $x_0 \not
\in \bigcap \{ B \in \mathbb{W}: \dia^{-1}A \subseteq B\}$, proving the inclusion and the lemma. \qed

\begin{cor}
$\Box^{-1}$ is an open operator in $ML$.
\end{cor}

\proof By the duality of $\dia^{-1}$ and $\Box^{-1}$.\qed

\begin{defi}
A formula $\phi \in ML^{+}$ is \emph{syntactically closed} if all occurrences of nominals and $\dia^{-1}$ in
$\phi$ are positive, and all occurrences of $\Box^{-1}$ in $\phi$ are negative; $\phi$ is \emph{syntactically
open} if all occurrences of $\dia^{-1}$ and nominals in $\phi$ are negative, and all occurrences of $\Box^{-1}$
in $\phi$ are positive. Clearly $\neg$ maps syntactically open formulae to syntactically closed formulae, and
vice versa.
\end{defi}

\begin{lem}
Every syntactically closed formula from $ML^{+}$ is a closed formula, and every syntactically open formula is an
open formula.\label{SyntacticallyClosedLemma}
\end{lem}

\proof By structural induction on syntactically open / closed formulae, written in negation normal form, using
the facts that $\dia$ and $\Box$ are open and closed operators, $\dia^{-1}$ is a closed operator, $\Box^{-1}$
and open operator and the fact that singletons are closed in descriptive frames. \qed

\begin{lem}
Let $\phi(q_1, \ldots, q_n, p, \nomi_1, \ldots \nomi_m)$ be a syntactically closed formula which is positive in
$p$ and with ${\sf PROP}(\phi) = \{q_1, \ldots, q_n, p \}$ and ${\sf NOM}(\phi) = \{\nomi_1, \ldots \nomi_m \}$.
Let $\mathfrak{F} = \langle W, R, \mathbb{W}\rangle$ be a descriptive frame in $ML$. Then for all $Q_1, \ldots, Q_n
\in \mathbb{W}$, $x_1, \ldots, x_m \in W$, and $C \in \mathbf{C}(\mathbb{W})$, it is the case that $\phi(Q_1,
\ldots, Q_n, C, \{x_1 \}, \ldots, \{x_m \})$ is closed in $T(\mathfrak{F})$. \label{ATypeClosedLemma}
\end{lem}

\proof We assume that $\phi$ is written in negation normal form, hence we may also assume that $\Box^{-1}$ does
not occur, as all occurrences have to be negative, and rewriting in negation normal form will change these into
$\dia^{-1}$. We proceed by induction on $\phi$. If $\phi$ is $\top$, $\bot$ or one of $q_1, \ldots, q_n, p, i_1,
\ldots i_m$ it is clear that $\phi(Q_1, \ldots, Q_n, \{x_1 \}, \ldots, \{x_m \}, C)$ is a closed set. This is
also the case if $\phi$ is the negation of a propositional variable from $q_1, \ldots, q_n$. The case when
$\phi$ is $\neg p$ does not occur.

The cases for $\land$ and $\lor$ follow since the finite unions and intersections of closed sets are closed. The
cases for $\dia$ and $\dia^{-1}$ follow from corollary \ref{diaClosedCor} and \ref{InvDiaClosedOperatorLemma},
respectively. The case for $\Box$ follows is immediate from the fact that $\Box$ is a closed operator, as was
noted earlier.\qed

\begin{lem}[Esakia's Lemma for Syntactically Closed Formulae] Let\\
$\phi(q_1, \ldots, q_n, p, \nomi_1, \ldots \nomi_m)$ be a syntactically closed formula with ${\sf PROP}(\phi) =
\{q_1, \ldots, q_n, p \}$ and ${\sf NOM}(\phi) = \{\nomi_1, \ldots \nomi_m \}$ which is positive in $p$. Let
$\mathfrak{F} = \langle W, R, \mathbb{W}\rangle$ be a descriptive frame in $ML$. Then for all $Q_1, \ldots, Q_n \in
\mathbb{W}$, $x_1, \ldots, x_m \in W$ and downwards directed family of closed sets $\{C_i : i \in I \}$ it is
the case that

$$\phi(Q_1, \ldots, Q_n, \bigcap_{i \in I}C_i, \{x_1 \}, \ldots,
\{x_m \}) = \bigcap_{i \in I} \phi(Q_1, \ldots, Q_n,  C_i, \{x_1
\}, \ldots, \{x_m \}).$$ \label{TypeAEsakiaLemma}
\end{lem}

\proof The proof is by induction on $\phi$. For brevity we will omit the parameters $Q_1, \ldots, Q_n$, $x_1,
\ldots, x_m$ when writing (sub)formulae. We assume that formulae are written in negation normal form, i.e we may
also assume that $\Box^{-1}$ does not occur, as all occurrences have to be negative, and rewriting in negation
normal form will change these into $\dia^{-1}$. The cases when $\phi$ is $\bot$, $\top$ or among $q_1, \ldots,
q_n, p, i_1, \ldots i_m$ are trivial, as are the cases when $\phi$ is the negation of a propositional variable
among $q_1, \ldots, q_n$. The inductive step in the case when $\phi$ is of the form $\gamma_1 \land \gamma_2$ is
also trivial.

Suppose $\phi$ is of the form $\gamma_1 \lor \gamma_2$. We have to show that $\gamma_1(\bigcap_{i \in I} C_i)
\cup \gamma_2(\bigcap_{i \in I} C_i) = \bigcap_{i \in I}(\gamma_1(C_i) \cup \gamma_2(C_i))$. The interesting
inclusion is from right to left, so assume that $x_0 \not \in \gamma_1(\bigcap_{i \in I} C_i) \cup
\gamma_2(\bigcap_{i \in I} C_i)$, i.e $x_0 \not \in \bigcap_{i \in I} \gamma_1( C_i) \cup \bigcap_{i \in I}
\gamma_2( C_i)$, by the induction hypothesis. Thus there exists $C_1, C_2 \in \{C_i : i \in I \}$ such that $x_0
\not \in \gamma_1(C_1)$ and $x_0 \not \in \gamma_2(C_2)$. By the downward directedness of $\{C_i : i \in I \}$
there is a $C \in \{C_i : i \in I \}$ such that $C \subseteq C_1 \cap C_2$. Thus, since $\gamma_1$ and
$\gamma_2$ are positive and hence upwards monotone in $p$, it follows that $x_0 \not \in \gamma_1(C)$ and $x_0
\not \in \gamma_2(C)$, and hence that $x_0 \not \in \bigcap_{i \in I}(\gamma_1(C_i) \cup \gamma_2(C_i))$.

Suppose $\phi$ is of the form $\dia \gamma$. We have to show that $\dia \gamma (\bigcap_{i \in I} C_i) =
\bigcap_{i \in I} \dia \gamma (C_i)$. By the inductive hypothesis we have $\dia \gamma (\bigcap_{i \in I} C_i) =
\dia \bigcap_{i \in I} \gamma (C_i)$. If $\gamma(C_i) = \emptyset$ for some $C_i$, then $\dia \bigcap_{i \in I}
\gamma (C_i) = \emptyset = \bigcap_{i \in I} \dia \gamma (C_i)$, so we may assume that $\gamma(C_i) \neq
\emptyset$ for all $i \in I$. Then, by Lemma \ref{ATypeClosedLemma}, $\{\gamma (C_i): i \in I\}$ is a family of
non-empty closed sets. Moreover, $\{\gamma (C_i): i \in I\}$ is downwards directed. For, consider any finite
number of members of $\{\gamma (C_i): i \in I\}$, $\gamma(C_1), \ldots, \gamma(C_n)$, say. Then there is a $C
\in \{C_i : i \in I\}$ such that $C \subseteq \bigcap_{i = 1}^{n} C_i$. But then $\gamma(C) \in \{\gamma (C_i):
i \in I\}$ and $ \gamma(C) \subseteq \bigcap_{i = 1}^{n} \gamma(C_i)$ by the upwards monotonicity of $\gamma$ in
$p$. Now we may apply Lemma \ref{DiamondEsakiaLemma} and conclude that $\dia \bigcap_{i \in I} \gamma (C_i) =
\bigcap_{i \in I} \dia \gamma C_i)$.

The case when $\phi$ is of the form $\dia^{-1} \gamma$ is verbatim the same the previous case, except that we
appeal to Lemma \ref{InvDiaEsakiaLemma} rather than Lemma \ref{DiamondEsakiaLemma} in the last step.

Lastly, consider the case when $\phi$ is of the form $\Box \gamma$. This follows by the inductive hypothesis and
the fact that $\Box$ distributes over arbitrary intersections of subsets of $W$.\qed

The  next lemma is needed for the following reason: for the proof
of canonicity of all formulae on which \SQEMA\ succeeds we will
need a version of Ackermann's lemma that is true of formulae in
the extended language $ML^{+}$, when interpreted over descriptive
frames. As already noted, the extension of an $ML^{+}$-formula
need in general not be an admissible set in such general frames.
This creates and obvious impediment for the proof of one direction
of the equivalence in the modal Ackermann's lemma, as the
interpretations of propositional variables \emph{must} be
admissible sets. We can push it through, however, at the price of
additional restrictions on formulae in terms of syntactic openness
and closedness.

\begin{lem}[Restricted Version of Ackermann's Lemma for Descriptive
Frames] Let:
\begin{itemize}
\item $\mathfrak{F} = \langle W, R, \mathbb{W} \rangle$ be a
descriptive or Kripke frame in $ML$;

\item $\!A(q_1,..., q_n, \nomi_1,..., \nomi_m)$ be a syntactically
closed formula with ${\sf PROP}(A) \subseteq \{q_1,..., q_n \}$ $\!$and
${\sf NOM}(A) \subseteq \{\nomi_1,...,\nomi_m \}$;

\item $\!B(q_1,..., q_n, p, \nomi_1,..., \nomi_m)$ be a syntactically
open formula with ${\sf PROP}(B)\!\subseteq\!\{q_1,..., q_n, p \}$ $\!$and
${\sf NOM}(B) \subseteq \{\nomi_1,..., \nomi_m \}$, which is downwards
monotone in $p$.
\end{itemize}

Then for all $Q_1, \ldots, Q_n \in \mathbb{W}$ and $x_1, \ldots, x_m \in W$:

$$B(Q_1, \ldots, Q_n, A(Q_1, \ldots, Q_n, \{x_1\}, \ldots,
\{x_m\}), \{x_1\}, \ldots, \{x_m\}) = W$$ if and only if there is
a $P \in \mathbb{W}$ such that

$$A(Q_1, \ldots, Q_n, \{x_1\}, \ldots, \{x_m\}) \subseteq P \textrm{
and } B(Q_1, \ldots, Q_n, P, \{x_1\}, \ldots, \{x_m\}) = W.$$
\label{RestrictedAckermann}
\end{lem}

\proof For the sake of brevity we will omit the parameters $Q_1, \ldots, Q_n, \{x_1\}, \ldots, \{x_m\}$ in what
follows, and simply write $A$, $B(P)$ etc. The implication from right to left follows by the downwards
monotonicity of $B$ in $p$. For the converse, if $\mathfrak{F}$ is a Kripke frame, i.e $\mathbb{W} = 2^{W}$, we can
simply take $P = A(Q_1, \ldots, Q_n, \{x_1\}, \ldots, \{x_m\})$, since all subsets of $W$ are admissible.

Now,  assume that $\mathfrak{F}$ is any descriptive frame and $B(A) = W$. Let $B'(p)$ be the negation of $B(p)$
written in negation normal form. Then $B'(p)$ is a syntactically closed formula, and $B'(A) = \emptyset$. We
need to find an admissible set $P \in \mathbb{W}$ such that $A \subseteq P$ and $B'(P) = \emptyset$. Since $A$
is a syntactically closed formula, it follows by Lemma \ref{SyntacticallyClosedLemma} that $A$ is a closed
subset of $W$ and hence that $A = \bigcap \{C \in \mathbb{W}: A \subseteq C\}$. Hence $\emptyset = B'(A) =
B'(\bigcap \{C \in \mathbb{W}: A \subseteq C\}) = \bigcap \{B'(C) : C \in \mathbb{W} \textrm{ and } A \subseteq
C\}$, by Lemma \ref{TypeAEsakiaLemma}. Again by Lemma \ref{SyntacticallyClosedLemma}, $\{ B'(C) : C \in
\mathbb{W}, A \subseteq C \}$ is a family of closed sets with empty intersection. Hence, by compactness, there
must be a finite subfamily, $C_1, \ldots, C_n$ say, such that $\bigcap_{i = 1}^{n} B'(C_i) = \emptyset$. But
then $C= \bigcap_{i = 1}^{n} C_i$ is an admissible set containing $A$, and $B'(C) = \emptyset$, i.e $B(C) = W$.
Hence we can choose $P = C$.\qed

We will refer to the equations which are pure formulae (i.e. do not contain propositional variables) as
\emph{pure equations}, and to the rest, as \emph{non-pure equations}.

\begin{lem}
During the entire (successful or unsuccessful) execution of
\SQEMA\ on any input formula from $ML$, all formulae on the
left-hand side of all non-pure \SQEMA-equations are syntactically
closed, and all formulae on the right-hand side of the non-pure
equations are syntactically open. \label{TypePreservationLemma}
\end{lem}
\proof We follow any one branch of the execution, proceeding by
induction. The starting system is of the form $\left \| \nomi \ra
\psi\right.$, where $\psi \in ML$. For this system the conditions
of the lemma hold, since $\nomi$ is syntactically closed and all
$ML$-formulae are both syntactically closed and open. Now suppose
that in the process of the execution we have reached a system
satisfying the conditions of the lemma, i.e. all formulae on the
left-hand side of all non-pure \SQEMA-equations are syntactically
closed, and all formulae on the right-hand side of the non-pure
equations are syntactically open. It is straightforward to check
that the application of any transformation rule to this system
will preserve these conditions. In the particular case when the
Ackermann-rule is applied, we note the following: (i) the pure
equations in the system contain no propositional variables, and
are hence disregarded in any application of the Ackermann-rule;
(ii) by the inductive hypothesis the disjunction of left-hand
sides which is substituted for the variable being eliminated are
syntactically closed; (iii) substituting a syntactically closed
formula for positive (resp., negative) occurrences of a variable
in a syntactically closed (resp., open) formula yields a
syntactically closed (resp., open) formula.\qed

The content of the next lemma is essentially that the global
satisfiability of systems of \SQEMA-equations is invariant under
all \SQEMA\ transformation rules, subject to the constraint that
$\nomi$ is evaluated to the current state.

\begin{lem}
Let $E_0, \ldots, E_r$ be the sequence of systems of equations
produced on one branch by \SQEMA\ when executed on a certain input
formula $\phi(q_1, \ldots, q_n) \in ML$ with ${\sf PROP}(\phi) =
\{q_1, \ldots, q_n \}$, and let $\nomi, \nomi_1, \ldots, \nomi_m$
be the nominals introduced during the execution. Further, let
$\phi_i(q_1, \ldots, q_n, \nomi, \nomi_1, \ldots, \nomi_m)$ be the
formula obtained by taking the conjunction of the equations in
$E_i$ for $0 \leq i \leq r$. Then for any descriptive or Kripke
frame $\mathfrak{F} = \langle W, R, \mathbb{W} \rangle$ and current
state $w \in W$, there are $Q_1, \ldots, Q_n \in \mathbb{W}$ and
$x_1, \ldots, x_m \in W$ such that

$$\phi_i(Q_1, \ldots, Q_n, \{ w \}, \{x_1\}, \ldots, \{x_m\}) =
W$$

if and only of there are $Q_1, \ldots, Q_n  \in \mathbb{W}$ and
$x_1, \ldots, x_m \in W$ such that

$$\phi_{i+1}(Q_1, \ldots, Q_n, \{ w \}, \{x_1\}, \ldots, \{x_m\}) =
W,$$

for $0 \leq i < r$. \label{SQEMAPreservationLemma}
\end{lem}

\proof Note that $\phi_1 = \nomi \ra \neg \phi$. For each $i$, the
system $E_{i+1}$ is obtained from the system $E_{i}$ by the
application of some transformation rule. Let $\mathfrak{F}$ be a
descriptive or Kripke frame, and $w$ a state in it. We need to
verify that, whichever transformation rule was applied, $\phi_{i}$
is globally satisfiable on $\mathfrak{F}$ if and only if $\phi_{i+1}$
is, subject to the constraint that $\nomi$ be always interpreted
as $w$. This is immediate to see for all the transformation rules,
except the Ackermann-rule, since they are based on simple
propositional and (modal and hybrid) semantic equivalences. The
only interesting case then, is the one for the Ackermann-rule.
Suppose $E_{i+1}$ is obtained from $E_{i}$ by application of this
rule. Then $\phi_i = \bigwedge_{j=1}^{l}(\alpha_j \ra p) \land
\bigwedge_{j=1}^{m}(\beta_j) \land \bigwedge_{j = 1}^{o}
\gamma_{j}$, where no $\alpha_j$ contains $p$, each $\beta_j$ is
negative in $p$, and no $\gamma_{j}$ contains any occurrence of
$p$. Note that all pure equations in the system $E_i$ will be
among the $\gamma_i$. Now all the $\alpha_j$'s are left-hand sides
of non-pure \SQEMA\ equations and hence, by lemma
\ref{TypePreservationLemma}, $\bigvee_{j = 1}^{l}\alpha_j$ is
syntactically closed. Further, if we eliminate the implication
sign from the $\beta_j$'s, each becomes a disjunction of the
negation of a left-hand side of a non-pure \SQEMA-equation with
the right-hand side of such an equation, which, again by lemma
\ref{TypePreservationLemma}, is syntactically open.

Then $\phi_{i+1} = \bigwedge_{j=1}^{m}(\beta'_j) \land \bigwedge_{j = 1}^{o} \gamma_{j}$, where each $\beta'_j$
is obtained from $\beta_j$ by substituting $\bigvee_{j = 1}^{l}\alpha_j$ for all occurrences of $p$. The proof
is complete once we appeal to lemma \ref{RestrictedAckermann}.\qed

\begin{thm}
If \SQEMA\ succeeds on a formula $\phi \in ML$, then $\phi$ is
locally d-persistent and hence canonical, and moreover the
first-order formula returned by \SQEMA\ is a local equivalent of
$\phi$.
\end{thm}

\proof Suppose that \SQEMA\ succeeds on $\phi \in ML$, and that
${\sf pure}(\phi)$ is the pure formula obtained as the conjunction
of the final system of \SQEMA-equations in the execution. Further,
for simplicity and without loss of generality, assume that the
execution does not branch because of disjunctions. We may make
this assumption since a conjunction of d-persistent formulae is
d-persistent, and the conjunction of local first-order
correspondents of modal formulae is a local first-order
correspondent for the conjunction of those formulae.

For the canonicity, let $\mathfrak{F} = \langle W, R, \mathbb{W} \rangle$ be a descriptive frame and $w \in W$.
Then:

$\mathfrak{F},w \not\forces \phi$ iff

for some model $\mathcal{M}$ based on $\mathfrak{F}$ with $\nomi$ denoting $w$, it is the case that $\mathcal{M}
\forces \nomi \ra \lnot \phi$.

Note that $\| \nomi \ra \lnot \phi$ is exactly the first system of
\SQEMA-equations obtained when the algorithm is run on $\phi$.
Hence ${\sf pure}(\phi)$ is obtained from $\nomi \ra \lnot \phi$
by the application of transformation rules.

Hence, $\mathcal{M} \forces \nomi \ra \lnot \phi$, for some model $\mathcal{M}$, based on $\mathfrak{F}$ with
$\nomi$ denoting $w$, iff

(by lemma \ref{SQEMAPreservationLemma} for descriptive frames) for some model $\mathcal{M}$, based on $\mathfrak{F}$
with $\nomi$ denoting $w$, it is the case that $\mathcal{M} \forces {\sf pure}(\phi)$, iff

for some model $\mathcal{M}$ based on the underlying Kripke frame $F$ with $\nomi$ denoting $w$, it is the case
that $\mathcal{M} \forces {\sf pure}(\phi)$, since we allow nominals to range over all singletons both in Kripke
and general frames.

Now by the Kripke frames version of Lemma \ref{SQEMAPreservationLemma}, and the fact that ${\sf pure}(\phi)$ is
obtained from $\nomi \ra \lnot \phi$ by the application of transformation rules, this is the case iff for some
model $\mathcal{M}$ based on the underlying Kripke frame $F$ with $\nomi$ denoting $w$, we have $\mathcal{M}
\forces \nomi \ra \lnot \phi$.

This, in turn, will be so iff $F, w \not\forces \phi$. This proves the local d-persistence of $\phi$.

Now, for the local first-order equivalence let $\mathfrak{F} = \langle
W, R \rangle$ be a Kripke frame and $w \in W$. As in the
formulation of lemma \ref{SQEMAPreservationLemma}, let $E_0,
\ldots, E_r$ be the sequence of systems of equations produced by
\SQEMA\ when executed on $\phi$ and let $\phi_j$ be the formula
obtained by taking the conjunction of the equations in $E_j$. We
define the \emph{translation of a system $E_j$}, ${\rm TR}(E_j)$,
to be the second order formula $\exists \overline{P} \exists
\overline{y} \forall x_0 {\rm ST}(\phi_j, x_0)$, where
$\overline{P}$ is the tuple of all predicate variables and
$\overline{y}$ the tuple of all variables corresponding to
nominals \emph{other than} $\nomi$, occurring in $\phi_j$. Note
that $y_i$, corresponding to $\nomi$, is the only free variable in
${\rm TR}(E_j)$, and that ${\rm TR}(E_r)$ is $\exists \overline{y}
\forall x_0 {\rm ST}({\sf pure}(\phi), x_0)$. Then:

$\mathfrak{F}, w \forces \phi$ iff

$\mathfrak{F} \models \forall \overline{P} ST(\phi, x_0)[x_0 := w ]$
iff

$\mathfrak{F} \models \forall \overline{P} \exists x_0  ST(\nomi \land
\phi, x_0)[y_i := w ]$ iff

$\mathfrak{F} \not \models \exists \overline{P} \forall x_0  ST(\nomi
\ra \neg \phi, x_0)[y_i := w ]$, i.e. iff

$\mathfrak{F} \not \models {\rm TR}(E_1)[y_i := w ]$.

Now, phrased in second-order logic, lemma
\ref{SQEMAPreservationLemma} says that
\newline $\mathfrak{F} \models {\rm TR}(E_j)[y_i := w ]$ if and only if
$\mathfrak{F} \models {\rm TR}(E_{j+1})[y_i := w ]$, for all $1 \leq j
< r$.

Hence we get that $\mathfrak{F}, w \forces \phi$ iff $\mathfrak{F} \not
\models \exists \overline{y} \forall x_0 {\rm ST}({\sf
pure}(\phi), x_0)[y_i := w]$,
\newline i.e that $\mathfrak{F}, w \forces \phi$ iff
$\mathfrak{F} \models \forall \overline{y} \exists x_0 \neg {\rm
ST}({\sf pure}(\phi),x_0)[y_i := w]$.

Hence $\forall \overline{y} \exists x_0 \neg {\rm ST}({\sf
pure}(\phi), x_0)$ is a local first-order correspondent for
$\phi$, and exactly what \SQEMA\ returns. Accordingly, $\forall
y_i \forall \overline{y} \exists x_0 \neg {\rm ST}({\sf
pure}(\phi), x_0)$ is a global first-order correspondent of
$\phi$.\qed

\section{Completeness results}

In this section we show that \SQEMA\ succeeds in computing the local first-order frame equivalent of every
Sahlqvist formulae, and then we extend that to every monadic inductive formula. While the latter result subsumes
the former, we nevertheless present both proofs in order to take the reader first through the simpler case.

\subsection{Completeness of \SQEMA\ on Sahlqvist Formulae}

Let us briefly recall what Sahlqvist formulae are (see
\cite{MLBook,vanBenthemBook,Sahlqvist}):

\begin{itemize}
\item a \emph{boxed atom} is a propositional variable, prefixed with finitely many (possibly none) boxes.

\item a \emph{Sahlqvist antecedent} is a formula built up from $\top$, $\bot$, boxed atoms, and
negative formulae, using $\land$, $\lor$ and diamonds.

\item a \emph{Sahlqvist implication} is a formula of the form $\phi \ra {\sf Pos}$ where $\phi$ is a Sahlqvist antecedent and ${\sf Pos}$
is a positive formula. In particular, note that any negative formula is a Sahlqvist antecedent.

\item a \emph{Sahlqvist formula} is built up from Sahlqvist implications
by applying conjunctions, disjunctions, and boxes.\footnote{The
definition in \cite{MLBook} requires that disjunctions are only
applied to formulae not sharing variables. Apparently, this
requirement is unnecessary.}

\end{itemize}

The following lemma is useful:

\begin{lem}
Let $\phi$ be a Sahlqvist formula, and $\phi'$ the formula obtained from $\neg \phi$ by importing the negation
over all connectives. Then $\phi'$ is a Sahlqvist antecedent. \label{SahlqvistAntecedentLemma}
\end{lem}

\proof Induction on the construction of $\phi$ from Sahlqvist implications. If $\phi$ is a Sahlqvist implication
$\alpha \ra {\sf Pos}$, negating and rewriting it as $\alpha \land \neg {\sf Pos}$ already turns it into a
Sahlqvist antecedent.

If $\phi = \Box \psi$, where $\psi$ satisfies the claim, then $\neg \phi \equiv \Diamond \neg \psi$ hence the
claim follows for $\phi$, because Sahlqvist antecedents are closed under diamonds.

Likewise, if $\phi = \psi_1 \land \psi_2$, where $\psi_1$ and $\psi_2$ satisfy the claim, then $\neg \phi \equiv
\neg \psi_1 \lor \neg \psi_2$ hence the claim follows for $\phi$, because Sahlqvist antecedents are closed under
disjunctions.

The case of $\phi = \psi_1 \lor \psi_2$ is completely analogous.\qed

\begin{cor}
Every Sahlqvist formula is semantically equivalent to a negated Sahlqvist antecedent, and hence to a Sahlqvist
implication.
\end{cor}

Sahlqvist's theorem \cite{Sahlqvist} states that all Sahlqvist formulae are elementary and canonical. For a
thorough discussion on Sahlqvist formulae and proof of Sahlqvist's theorem, see e.g. \cite{MLBook}.

\begin{lem}
Let $E$ be a system of \SQEMA\ equations of the form $\nomj \ra \beta$, with $\nomj$ a nominal and $\beta$ a
Sahlqvist antecedent built up without using $\lor$, except possibly inside negative formulae. Let $p$ be any
propositional variable occurring both negatively and positively in $E$. Then $E$ can be transformed, using only
the $\land$-rule, $\dia$-rule, $\Box$-rule and Ackermann-rule, into a system $E'$, again with all equations of
the form $\nomj \ra \beta$, which does not contain $p$. \label{SahlqvistEliminationLemma}
\end{lem}

\proof All positive occurrences of $p$ are in boxed atoms, occurring at most in the scope of conjunctions and
diamonds. We first have to separate $p$, i.e. transform $E$ into a system in which the all equations in which
$p$ occurs positively are of the form $\gamma \ra p$, with $p$ not occurring in $\gamma$. Start by applying the
$\dia$-rule and $\land$-rule to separate the boxed atoms of $p$. The system then obtained is still one in which
all equations are of the form $\nomj \ra \beta$, with $\nomj$ a nominal and $\beta$ a Sahlqvist antecedent ---
recall that pure formulae are regarded as both positive and negative. All equations in which $p$ occurs
positively are now of the form $\nomj \ra \Box^{n} p$. By applying the $\Box$-rule these are transformed into
equations of the form $(\dia^{-1})^{n} \nomj \ra p$. Note that the system is now no longer in the desired form
of implications from nominals to Sahlqvist antecedents. This is remedied by applying the Ackermann-rule to
eliminate $p$, together with all equations of the form $(\dia^{-1})^{n} \nomj \ra p$. Note that pure formulae
are substituted for negative occurrences of $p$, hence the substitution leaves negative formulae negative, and
hence Sahlqvist antecedents as Sahlqvist antecedents. The system now obtained is once again of the desired form,
and $p$ has been eliminated.\qed

\begin{thm}
\SQEMA\ succeeds on every Sahlqvist formula.
\end{thm}

\proof Let $\phi$ be a Sahlqvist formula. Let us see what \SQEMA\ does with it. In step one, $\neg \phi$ is
formed and the negation imported over all connectives. Call the formula so obtained $\phi'$. By Lemma
\ref{SahlqvistAntecedentLemma} $\phi'$ is a Sahlqvist antecedent. $\phi'$ is now transformed into the form
$\bigvee_{j = 1}^{n} \alpha_j$ with each $\alpha_j$ a Sahlqvist antecedent built up from $\top$, $\bot$, boxed
atoms and negative formulae without using disjunctions by distributing conjunctions and diamonds over
disjunctions. Note that it is possible to ``extract'' all disjunctions, except possibly some within negative
formulae, since in $\phi'$ none of these disjunctions occur within the scope of boxes.

From here the algorithm proceeds on each disjunct separately; we will follow one of them, $\alpha_j$. Let $p_1,
\ldots, p_m$ be an arbitrary ordering of the variables occurring in $\alpha_j$. The initial system of equations
is
$$\left \| \begin{array}[l]{l} \nomi \ra \alpha_j.
\end{array} \right.
$$

Note that this is a system of the form required by Lemma \ref{SahlqvistEliminationLemma}. If $p_1$ occurs only
positively (negatively) \SQEMA\ eliminates it by substituting it with $\top$ ($\bot$). If $p_1$ occurs both
negatively and positively, it follows from Lemma \ref{SahlqvistEliminationLemma} that \SQEMA\ will eliminate it.
In both cases the resulting system is again one of the form required by Lemma \ref{SahlqvistEliminationLemma}.
Propositional variables $p_2, \ldots, p_m$ remain to be eliminated. Proceeding inductively, noting that after
each elimination Lemma \ref{SahlqvistEliminationLemma} remains applicable, it follows that \SQEMA\ will succeed
in doing this. Negating and translating the resulting pure formula, \SQEMA\ obtains the first-order (local)
frame equivalent for $\alpha_j$. Taking the conjunction of these equivalents for $1 \leq j \leq n$ yields the
first-order (local) frame equivalent for the Sahlqvist formula $\phi$ given as input. \qed

\subsection{Completeness of \SQEMA\ on Monadic Inductive Formulae}

\begin{defi}
Let $\sharp$ be a symbol not belonging to $\mathrm{ML}^{+}$. Then a \emph{box-form of $\sharp$} in
$\mathrm{ML}^+$ is defined recursively as follows:
\begin{enumerate}
\item $\sharp$ is a box-form of $\sharp$;
\item If $\mathbf{B}(\sharp)$ is a box-form of $\sharp$ and $\Box$ is a box-modality then $\Box \mathbf{B}(\sharp)$ is
a box-form of $\sharp$.
\item If $\mathbf{B}(\sharp)$ is a box-form of $\sharp$ and $A$ is a positive formula then $A \ra \mathbf{B}(\sharp)$
is a box-form of $\sharp$.
\end{enumerate}
\end{defi}

Thus, box-forms of $\#$ are, up to semantic equivalence, of the type
\[A_{0}\rightarrow \square _{1}(A_{1}\rightarrow
\ldots \square_{n}(A_{n}\rightarrow \#)\ldots ) \]

where $\square _{1},\ldots ,\square _{n}$  are compositions of boxes in $\mathrm{ML}^+$ and $A_{1},\ldots
,A_{n}$ are positive formulae (possibly, just $\top$).

\begin{defi}
By substituting a propositional variable $p$ for $\sharp$ in a box-form $\mathbf{B}(\sharp)$ we obtain a
\emph{box-formula of $p$}, namely $\mathbf{B}(p)$. The last occurrence of the variable $p$ is the \emph{head} of
$\mathbf{B}(p)$ and every other occurrence of a variable in $\mathbf{B}(p)$ is \emph{inessential} there.
\end{defi}

\begin{defi}[\cite{GorankoVakarelov3,CGVAiml5}]
A \emph{monadic regular formula} is a modal formula built up from $\top$, $\bot$, positive formulae and negated
box-formulae by applying conjunctions, disjunctions, and boxes.
\end{defi}

\begin{defi}
The \emph{dependency digraph} of a set of box-formulae $\mathcal{B}=\{\mathbf{B}_{1}\mathbf{(}p_{1}),\ldots ,
\mathbf{B}_{n}\mathbf{(}p_{n})\}$ is a digraph $G_{\mathcal{B}}=\left\langle V,E\right\rangle $ where $V
=\{p_{1},\ldots ,p_{n}\}$ is the set of heads in $\mathcal{B}$, and edge set $E$, such that $p_{i} E p_{j}$ iff
$p_{i}$ occurs as an inessential variable in a box from $\mathcal{B}$ with a head $p_{j}$. A digraph is
\emph{acyclic} if it does not contain oriented cycles (including loops).

The dependency digraph of a formula is the dependency digraph of the set of box-formulae that occur as
subformulae of that formula.
\end{defi}

\begin{defi}[\cite{GorankoVakarelov3,CGVAiml5}]
A \emph{monadic inductive formula (MIF)} is a monadic regular formula with an acyclic dependency digraph.
\end{defi}

\begin{exa}
An example of a monadic inductive formula, which is not a Sahlqvist formula (see
\cite{GorankoVakarelov1,GorankoVakarelov3}), is:
\[
D=p\wedge \square (\dia p\rightarrow \square q)\rightarrow \dia \square \square q\equiv \lnot p\vee \lnot
\square (\dia p\rightarrow \square q)\vee \dia \square \square q,
\]
obtained as a disjunction of the negated box-formulae $\lnot p$ and $\lnot \square (\dia p\rightarrow \square
q)$, and the positive formula $\dia \square \square q$. The dependency digraph of $D$ over the set of heads
$\{p,q\}$ has only one edge, from $p$ to $q$.

An example of a regular, but non-inductive formula is $\Box((\neg \Box p \lor q) \lor (\neg q \lor \Box p)) \lor
\neg p$, because the heads $p$ and $q$ depend on each other.
\end{exa}

The abbreviation ${\sf NegMIF}$ will be used for the negation of a monadic inductive formula in negation normal
form. Note that the class of ${\sf NegMIF}$s in a language are precisely those formulae built from $\top$,
$\bot$, negative formulae and box-formulae using conjunction, disjunction and diamonds, which have acyclic
dependency digraphs. The abbreviation ${\sf NegMIF}^*$ will be used for ${\sf NegMIF}$s built up without the use
of disjunction, i.e. the class of all formulae of the language built from $\top$, $\bot$, negative formulae and
box-formulae using conjunctions and diamonds, which have acyclic dependency digraphs.

We will call a system of \SQEMA\ equations a \emph{${\sf NegMIF}$-system} (\emph{${\sf NegMIF}^*$-system}) if it
has the form

$$\left \| \begin{array}[l]{l}
\mathbf{i}_1 \ra \alpha_1\\
\vdots\\
\mathbf{i}_n \ra \alpha_n\\
\end{array} \right.,
$$

where each $\mathbf{i}_i$ is a nominal, and the formula $\alpha_1 \land \ldots \land \alpha_n$, obtained by
taking the conjunction of all consequents of equations in the system, is a ${\sf NegMIF}$ (${\sf NegMIF}^*$) in
$\mathrm{ML}^+$.

\begin{lem}
Let $E$ be a ${\sf NegMIF}^*$-system of \SQEMA-equations, and $p$ any propositional variable occurring in $E$.
Then $p$ can be eliminated from $E$, either by substitution with $\top$ or $\bot$, or by application of the
\SQEMA\ transformation rules. Moreover, the system of equations obtained after the elimination of $p$ will again
be a ${\sf NegMIF}^*$-system.\label{NegMIFStarLemma}
\end{lem}

\proof Suppose $p$ occurs only positively (negatively) in the system, then we substitute $\top$ ($\bot$) for all
its occurrences, eliminating it, and yielding a  ${\sf NegMIF}^*$-system.

So suppose $p$ occurs both positively and negatively. We will separate out all positive occurrences to prepare
for the application of the Ackermann-rule. The only positive occurrences of $p$ are heads of (possibly trivial)
box-formulae in the consequents of the equations of the system. Let $\mathbf{i}_{i} \ra \alpha$ be such an
equation, where $\alpha$ contains positive occurrences of $p$. Exhaustive application of the $\land$-rule and
the $\dia$-rule splits $\mathbf{i}_{i} \ra \alpha$ into equations of the forms $\mathbf{j} \ra \dia \mathbf{k}$,
$\mathbf{l} \ra {\sf Box}$ and $\mathbf{m} \ra {\sf Neg}$, with $\mathbf{j}, \mathbf{k}, \mathbf{l}, \mathbf{m}$
nominals, ${\sf Box}$ a box-formula and ${\sf Neg}$ a negative formula. As all box-formulae have been left
intact, the dependency digraph is unchanged. Hence, after the application of these rules we still have a ${\sf
NegMIF}^*$-system.

Now it remains to separate the positive occurrences of $p$ out of the equations of the form $\mathbf{l} \ra {\sf
Box}$, where ${\sf Box}$ is of the form $\Box_1 (A_1 \lor \Box_2(\ldots \Box_n(A_n \lor p )\ldots ))$ where each
$A_i$ is a negative formula and each $\Box_i$ a finite, possibly empty, sequence of box modalities. Successive
alternative applications of the Left shift $\lor$-rule and Left shift $\Box$-rule transforms the equation
$$\mathbf{l} \ra A_0 \lor \Box_1 (A_1 \lor \ldots \Box_n(A_n \lor p )\ldots
)) $$

into
$$\neg A_n \land \dia^{-1}_{n}(\ldots \dia^{-1}_{1}( \mathbf{l} \land \neg
A_0)\ldots) \ra p.$$

The antecedent in the above equation is a positive formula, not containing $p$, because the dependency graph is
loopless. Hence all positive occurrences of $p$ in the system now occur as the consequents in equations of the
form ${\sf Pos} \ra p$, where ${\sf Pos}$ is a positive formula not containing $p$. Let $\rho$ be the
disjunction of all the antecedents of the equations of the form ${\sf Pos} \ra p$. Then, by applying the
Ackermann-rule, all equations of this form are deleted and $p$ is eliminated by substitution of the positive
formula $\rho$ for every negative occurrence of $p$. Thus, the resulting system does not contain $p$, all
antecedents of equations are nominals, and all consequents of equations are built up from negative formulae and
box-formulae, by using only conjunctions and diamonds.

To show that the resulting system is again ${\sf NegMIF}^*$, it remains to show that the dependency digraph is
acyclic. We will do so by showing that whenever a new arc $(q,u)$ was introduced by the application of the
Ackermann-rule, there was already a directed path from vertex $q$ to vertex $u$ in the digraph \emph{before} the
substitution. Indeed, for the only way $(q,u)$ could have been introduced was by the substitution of $\rho$ for
an inessential occurrence of $p$ in some box-formula with $u$ as head. But then $q$ must occur in $\rho$, hence,
by the construction of $\rho$, it must have occurred inessentially in some box-formula headed by $p$. But then
$(q,p)$ and $(p,u)$ were arcs in the dependency digraph before the application of the Ackermann-rule, giving the
desired path. Thus, the application of the Ackermann-rule cannot introduce cycles in a previously acyclic
dependency graph, which completes the argument. \qed

\begin{thm}
\SQEMA\ succeeds on all conjunctions of monadic inductive formulae.
\end{thm}

\proof An easy inductive argument, noting that the initial \SQEMA-systems for every conjunction of monadic
inductive formulae are ${\sf NegMIF}^*$-systems, and using Lemma \ref{NegMIFStarLemma}.\qed

\begin{rem}
It has been proved in \cite{GorankoVakarelov1} (see also \cite{GorankoVakarelov3}) that every Sahlqvist formula
is tautologically equivalent to a conjunction of inductive formulae, which now provides another proof of
completeness of \SQEMA\ for Sahlqvist formulae.
\end{rem}

\begin{cor}[Sahlqvist theorem for inductive formulae, see \cite{GorankoVakarelov3}] All monadic inductive formulae are
elementary and canonical.
\end{cor}

\begin{conj}
All modal formulae on which \textsf{SQEMA}\ succeeds are locally equivalent to inductive formulae.
\end{conj}

\section{Concluding remarks and further work}
\label{ConcludingRemarks}

The algorithm \SQEMA\ is presented here in its core version, which
is apparently already quite powerful. In particular, we are
currently not aware of examples whereby \SQEMA\ fails on a modal
formula on which the current implementations of either SCAN or DLS
succeed.

Furthermore, \SQEMA\ is amenable to various extensions, some of
which are being developed in forthcoming sequels
\cite{CGVsqema2,CGVsqema3,CGVsqema4} to this paper. In particular:

\begin{description}
\item[Monotonicity-based version]

While the condition on the formula $B(P)$ in the formulation of
Ackermann's lemma is stated in syntactic terms, the proof of that
lemma uses a \emph{semantic} property of $B(P)$, viz
\emph{upwards, resp. downwards monotonicity} with respect to $P$.
Thus, the lemma can be strengthened accordingly to a
\emph{strong}, or \emph{semantic} Ackermann's lemma, which
accordingly yields a stronger, semantic version of the respective
Ackermann rule used by \SQEMA. Moreover, testing a modal formula
for monotonicity in a propositional variable occurring in it is
decidable, which enables the algorithmic use of such rule in a
stronger version of \SQEMA, albeit at the price of possibly higher
complexity.

\item[Polyadic languages] The algorithm can be accordingly extended to work in arbitrary
polyadic modal languages (presented as \emph{purely modal
languages} in \cite{GorankoVakarelov1,GorankoVakarelov3}) and to
succeed on all polyadic inductive formulae (ibid.).

\item[Complex formulae] Furthermore, it can be extended with suitable rules to succeed on
the so called \emph{complex formulae} introduced in
\cite{Vakarelov1}. These can be converted to inductive formulae by
using rather non-trivial substitutions.

\item[Least fixed points] As shown in \cite{GorankoVakarelov3}, all regular formulae (in
arbitrary polyadic languages) have equivalents in first-order
logic extended with least fixed points FO+LFP. These equivalents
can be computed by a suitable extension of \textsf{SQEMA}\ with a
recursive version of the Ackermann-rule. For instance, that
extension succeeds on the G\"{o}del-L\"{o}b formula \ $\square
(\square p\rightarrow p)\rightarrow \square p$ and on Segerberg's
induction axiom $IND=[2](q\rightarrow \lbrack 1]q)\rightarrow
(q\rightarrow \lbrack 2]q)$. Note that these formulae are not
canonical anymore.

\item[Theory reasoning] The algorithm \textsf{SQEMA}\ is amenable to further extensions
with rules capturing \emph{theory reasoning}, which enables
computing first-order correspondents relativized over classes of
frames, and thus succeeding on notorious cases of elementary
canonical formulae, such as modal reduction principles over the
class of transitive frames.

\end{description}

\section*{Acknowledgements}

This work was mostly done while V. Goranko was on a sabbatical
leave from Rand Afrikaans University. He would also like to
acknowledge the financial support for his research from the
National Research Foundation of South Africa. Some of the work on
this paper was carried out during the visit of V. Goranko and W.
Conradie to the University of Manchester, organized by  Renate
Schmidt and funded by the British Research Council, which we
gratefully acknowledge. The cooperation of Dimiter Vakarelov was
supported by EC COST Action 274 ``Theory and Applications of
Relational Structures as Knowledge Instruments'' (TARSKI), {\tt
http:$\backslash\backslash$www.tarski.org}, and by the project No
1510 of the Bulgarian Ministry of Science and Education. We would
also like to thank Renate Schmidt and Rajeev Gore for useful
discussions and remarks, as well as the referees for some
corrections and constructive criticism.

\end{document}